\documentclass[12pt,letterpaper]{article}
\pdfoutput=1
\usepackage{amssymb}
\usepackage{amsmath}
\usepackage{amsfonts}
\usepackage{epsfig}
\usepackage{color}
\usepackage{hyperref}
\usepackage{cite}

\usepackage{bm}
\outer\def\beginsection#1\par{\medbreak\bigskip
      \message{#1}\leftline{\bf#1}\nobreak\medskip
\vskip-\parskip
      \noindent}

\def\m{\mu}

\def\f{\varphi}
\def\a{\alpha}

\def\S{\Sigma}

\def\Ord{{\cal O}}

\def\q{q}
\def\be{\begin{equation}}
\def\ee{\end{equation}}
\def\ba{\begin{eqnarray}}

\newcommand{\Ered}[1]{{\color{black} #1}}

\newcommand{\ea}[1]{\begin{align} #1 \end{align}}

\newcommand{\seal}[2]{\begin{subequations}\label{#1} \begin{align} #2 \end{align}\end{subequations}}
\newcommand{\seq}[1]{\begin{equation} \begin{split} #1 \end{split} \end{equation}}

\newcommand{\LN}[2]{\log | z_{#1}-z_{#2} |}
\newcommand{\zz}[2]{z_{#1}-z_{#2}}
\newcommand{\zbzb}[2]{\bar{z}_{#1}-\bar{z}_{#2}}
\newcommand{\aqk}[1]{ \alpha' q k_{#1}}
\newcommand{\qk}[1]{ q k_{#1}}
\newcommand{\teq}[1]{\f_{#1}\epsilon_{#1} q}

\newcommand{\tbeq}[1]{\bar{\f_{#1}}\bar{\epsilon}_{#1} q}

\newcommand{\te}[2]{(\f_{#2}\epsilon_{#2}^{#1})}

\newcommand{\tbe}[2]{(\bar{\f_{#2}}\bar{\epsilon}_{#2}^{#1})}

\def\bA{\bar{A}}
\def\bC{\bar{C}}
\def\M{\mathcal{M}}
\def\S{\mathcal{S}}

\usepackage[titles]{tocloft}

\usepackage{titlesec}
\titleformat*{\section}{\large  \bfseries }
\titleformat*{\subsection}{\normalsize  \bfseries }

\begin{document}

\begin{titlepage}
\hfill \hbox{NORDITA-2017-054}
\vskip 1.5cm
\begin{center}
{\Large \bf
The B-field soft theorem and its unification with the graviton and dilaton
}
 
\vskip 1.0cm {\large Paolo
Di Vecchia$^{a,b}$,
Raffaele Marotta$^{c}$, Matin Mojaza$^{d}$
} \\[0.7cm] 
{\it $^a$ The Niels Bohr Institute, University of Copenhagen, Blegdamsvej 17, \\
DK-2100 Copenhagen \O , Denmark}\\
{\it $^b$ Nordita, KTH Royal Institute of Technology and Stockholm 
University, \\Roslagstullsbacken 23, SE-10691 Stockholm, Sweden}\\[2mm]
{\it $^c$  Istituto Nazionale di Fisica Nucleare, Sezione di Napoli, Complesso \\
 Universitario di Monte S. Angelo ed. 6, via Cintia, 80126, Napoli, Italy}\\[2mm]
 {\it $^d$  Max-Planck-Institut f\"ur Gravitationsphysik, \\
Albert-Einstein-Institut, Am M\"uhlenberg 1, 14476 Potsdam, Germany}
\end{center}
\begin{abstract}
In theories of Einstein gravity coupled with a dilaton and a two-form,
a soft theorem for the two-form, known as the Kalb-Ramond B-field, has so far been missing. 
In this work we fill the gap, and in turn formulate a unified soft theorem valid
for {gravitons, dilatons and B-fields} in any tree-level 
scattering amplitude involving the three {massless} states.
The new soft theorem is fixed by means of on-shell gauge invariance 
and enters at the subleading order of the graviton's soft theorem. 
In contrast to the subsubleading soft behavior of gravitons and dilatons, 
we show that the soft 
behavior of B-fields at this order cannot be fully fixed by gauge invariance. 
Nevertheless, we show that it is possible   to establish a gauge invariant 
decomposition  of the amplitudes to any order in the soft expansion.
We check explicitly the new soft theorem in the bosonic
string and in Type II superstring theories, and furthermore demonstrate that, 
at the next order in the soft expansion, totally gauge invariant terms appear in both string theories
 which cannot be factorized into a soft theorem.

\end{abstract}
\end{titlepage}

\tableofcontents

\vspace{5mm}

\section{Introduction}
\label{intro}

{There has been a huge effort in the last few years to connect the soft
 behavior of  scattering amplitudes 
 of massless particles to  underlying symmetries of 
the theory. This connection has been made explicit  in  the case of 
gauge, gravity and higher-spin theories, where the soft factorization theorems of the scattering amplitudes
through subleading orders have been shown to follow from their on-shell gauge invariance~\cite{Broedel:1406,BDDN,DMM3,DiVecchia:2015jaq,Roiban:2017iqg}. 
Their relation
to asymptotic symmetries were recently suggested~\cite{Strominger:2013lka,Strominger:2013jfa}, and are now being being vastly explored (see e.g. the recent review~\cite{Strominger:1703} and references therein, as well as the more recent paper~\cite{Campiglia:1703}, which discusses also scalar soft theorems from asymptotic symmetries).}

Similar results have, on the other hand, also recently been obtained 
in theories where global  
internal\cite{ArkaniHamed:2008gz,Kampf:2013vha,Low:1412,Du:2015esa} or
global space-time symmetries\cite{Boels:2015pta,Huang:2015sla,DiVecchia:2015jaq,DiVecchia:1705,Bianchi:2016viy,Guerrieri:1705} are 
spontaneously broken. 
In these cases,  it is the spontaneously broken symmetry that determines
the soft behavior of amplitudes with soft Nambu-Goldstone bosons.

The focus of this paper concerns the gravitational S-matrix
in theories of gravity coupled with a dilaton and a two-form.
It is well known from string theory that amplitudes of this theory can be described in a unified way, which we in short will describe. Nevertheless, while it is known that the graviton and dilaton obey soft theorems,
a soft theorem for the two form is still missing~\cite{DMM1,Cheung:2017ems}.
The aim  of this paper is to 
fill this gap, and to derive a unified soft theorem valid universally for the 
graviton, dilaton and the two-form.

The two-form appears particularly in theories of supergravity and string 
theory, where it is known as the Kalb-Ramond two-form B-field. In string
theory it enters as a closed string massless state accompanying the graviton and dilaton. {Their soft behavior, and particularly the soft theorems obeyed by the string graviton and dilaton were recently derived in Refs.~\cite{Schwab:2014sla,Bianchi:1512,DMM1,DMM2, DMM3, Sen:1702,Sen:1703,Sen:1706}, 
extending the well-known graviton~\cite{Weinberg:1964,Weinberg:1965,Gross:1968,Jackiw:1968} and less-known 
dilaton~\cite{SoftDilaton,Shapiro:1975cz} soft theorems from the 60s and 70s to one higher 
order.  (See also Ref.~\cite{BianchiR2phi,Schwab:2014fia, Bianchi:2015yta,
DiVecchia:2015bfa,Guerrieri, DiVecchia:2015srk} for other string theory related soft theorems.)}

The B-field also appears in 
gravity as a double-copy Yang-Mills theory. 
An easy way to understand this, which will later be useful, is by 
decomposing the double-copied Yang-Mills field into its constituent fields:
\ea{
A_\mu(k) \tilde{A}_\nu(k) &= 
\left [\frac{A_\mu \tilde{A}_\nu + A_\nu \tilde{A}_\mu}{2} 
- \frac{\epsilon_{\mu\nu}^\perp}{\sqrt{D-2}} \, A \cdot \tilde{A}
\right ] 
+ \left[ \frac{\epsilon_{\mu\nu}^\perp}{\sqrt{D-2}} \,  
A \cdot \tilde{A}
\right ] 
+\left [\frac{A_\mu \tilde{A}_\nu - A_\nu \tilde{A}_\mu}{2} 
\right ] 
\nonumber \\
&=g_{\mu \nu}(k) + \frac{\epsilon_{\mu\nu}^\perp}{\sqrt{D-2}}  \phi(k) + B_{\mu \nu}(k)
}
where $D$ is the number of spacetime dimensions and
\ea{
\epsilon_{\mu\nu}^\perp = \frac{\eta_{\mu \nu} - k_\mu \bar{k}_\nu-k_\nu \bar{k}_\mu}{\sqrt{D-2}}
\, , \quad
k^2 = \bar{k}^2=0 \, , \quad k\cdot \bar{k} = 1
}
such that $\eta^{\mu \nu} \epsilon_{\mu \nu}^\perp = \sqrt{D-2}$, and
$\epsilon_{\mu \nu}^\perp {\epsilon^\perp}^{\mu \nu} = 1$, while
$\bar{k}$ is an unphysical reference momentum.
The fields $g_{\mu \nu}$, $\phi$, and $B_{\mu \nu}$ are naturally identified with the gravitational field, the dilaton and the antisymmetric B-field, while $\epsilon^\perp_{\mu \nu}$ can be thought of as the dilaton `polarization tensor' (or dilaton projector).
{By the Kawai-Lewellen-Tye~\cite{Kawai:1985xq} and Bern-Carrasco-Johansson~\cite{Bern:1004} relations, double-copied amplitudes of Yang-Mills theory can be identified with amplitudes of gravity coupled with the dilaton and B-field, providing a unified description for amplitudes involving any of the three fields.}

From this double-copy 
construction one may na\"ively expect that also the B-field should 
obey soft theorems at the same order. After all, one may compute the 
amplitudes generically, and only in the end project the external states 
appropriately. However, the na\"ive expectation turns out not to hold, and
we will explain why.

\begin{figure}[tb]
\begin{center}
\includegraphics[width=.9\textwidth]{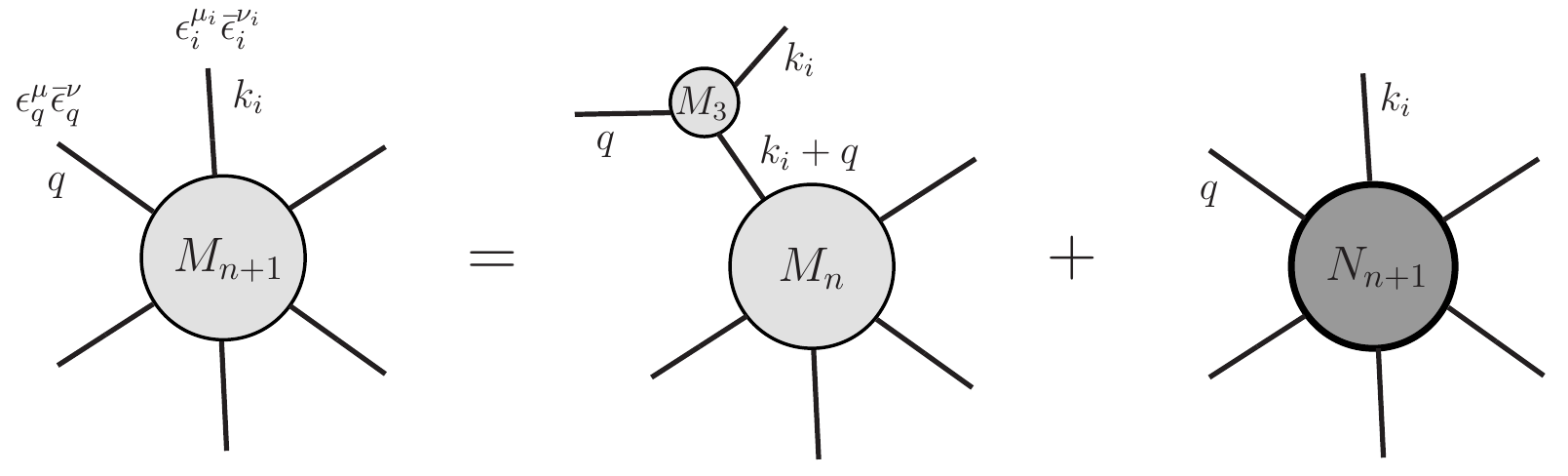}
\caption{
{Decomposition} of an $n+1$-point amplitude into 
a factorizing set, involving an exchange of a particle between the three-point 
amplitude $M_3$ and the $n$-point amplitude $M_n$, and the reminder set 
of diagrams $N_{n+1}$, which excludes factorization through the former 
channel.
}
\label{factorization}
\end{center}
\end{figure}

It is useful to first review how the soft theorems for the graviton and
dilaton can be derived by using gauge invariance of the amplitude.
Considering an {($n+1$)-point} amplitude involving double-copied Yang-Mills
states, i.e. gravitons, dilatons, and B-fields,
it is possible to decompose it into two
contributions, as
 depicted in Fig.~\ref{factorization}.
The decomposition of the amplitude as given in Fig.~\ref{factorization} can 
be written as:
\ea{
M_{n+1} = \sum_{i=1}^n M_3(q; k_i) \frac{1}{(k_i+q)^2}
M_n(k_i+q) + N_{n+1} (q; k_i) \, ,
\label{decomposition}
}
where dependence on all other $k_j \neq k_i$ is implicit.
We are giving a special role to the state carrying momentum $q$, since we will 
consider its soft behavior.
By using the relation
\mbox{$M_n = \epsilon_1^{\mu_1} \bar{\epsilon}_1^{\nu_1} \cdots \epsilon_n^{\mu_n} \bar{\epsilon}_n^{\nu_n} M_{n\, \mu_1 \nu_1 \ldots \mu_n \nu_n}$}, 
{the three-point amplitude entering in the above expression
can be rewritten in terms of differential operators acting on $M_n$}, given by~\cite{BDDN,DMM3}:
\ea{
M_3\left(q, k_i, - (k_i+q)\right ) {= 
2 \kappa_D} \epsilon_q^\mu \bar{\epsilon}_q^\nu \big[ k_{i\mu} -  i
q^\rho S_{i \, \mu \rho} \big ]\big [
 k_{i\nu} -  i
q^\sigma {\bar{S}}_{i\, \nu \sigma} \big ] \, ,
}
{where $\kappa_D$  is related to Newton's constant by 
$\kappa_D = \sqrt{8 \pi G_N^{(D)}}$}
and where
\ea{
S_{i\, \mu \rho} = i \left(\epsilon_{i \mu}
\frac{\partial}{\partial \epsilon_i^{\rho}} - \epsilon_{i \rho}
\frac{\partial}{\partial \epsilon_i^{\mu}} \right)
~;~{\bar{S}}_{i\, \nu \sigma} = i
\left({\bar{\epsilon}}_{i \nu}
\frac{\partial}{\partial {\bar{\epsilon}}_i^{\sigma}} - {\bar{\epsilon}}_{i \sigma }
\frac{\partial}{\partial {\bar{\epsilon}}_i^{\nu}} \right) \, .
}

The two 
contributions in Eq.~(\ref{decomposition}) are not independently 
gauge invariant.
We may therefore use gauge invariance to put certain constraints on the remainder function {$N_{n+1}$}.  
Considering the kinematical region where $q \ll k_i$ for any $i$, 
Eq.~\eqref{decomposition} can be expanded in $q$.
At tree-level, {$N_{n+1}$} does not have any poles in $q$ and hence at leading order one recovers Weinberg's soft theorem (loops do not modify this leading order result, see e.g.~\cite{Bern:2014oka,Broedel:2014bza} for a recent discussion):
\ea{
M_{n+1} = {\kappa_D} \epsilon_{q\,\mu} \bar{\epsilon}_{q\, \nu} \sum_{i=1}^n
\frac{k_i^\mu k_i^\nu}{k_i\cdot q}
M_n(k_i) + \Ord(q^0)
\label{lead}
}
{To project this expression on the physical states, one takes $\epsilon_{q\,\mu} \bar{\epsilon}_{q\, \nu} = \epsilon_{q\, \mu \nu}$ to be the polarization tensor of either the graviton, dilaton or B-field.
If the soft state is projected onto the B-field, whose polarization tensor is antisymmetric, this leading expression clearly vanishes}, but also if it is projected onto the dilaton, since ${\epsilon_{q\,\mu \nu}^{\perp} }\, k_i^\m k_i^\nu = k_i^2 + \Ord(q) = 0 + \Ord(q)$.
Only the graviton has a leading nonzero singular soft behavior. The expression is gauge invariant due to momentum conservation $\sum_{i=1} k_i^\mu = 0 + \Ord(q)$.

At every other order in the soft expansion, the two 
types of contributions  are related by gauge invariance; that is, by
\ea{
q_\mu \bar{\epsilon}_\nu M_{n+1}^{\mu \nu} = q_\nu \epsilon_\mu M_{n+1}^{\mu \nu} = 0 \, .
}
These two conditions are sufficient to fix 
completely the 
orders $q^0$ and $q^1$ in the soft expansion  of the 
amplitude, when $M_{n+1}^{\mu \nu}$ is symmetric in its indices
$\mu \nu$; i.e. when the soft state is either a graviton or a
dilaton~\cite{BDDN,DMM3}. 
We will in this work additionally show that if the soft state is a B-field, i.e. 
if $M_{n+1}^{\mu \nu}$ is antisymmetric in $\mu \nu$, this construction is sufficient to
fix completely its  order $q^0$  soft behavior, with the result being:
\ea{
M_{n+1} = -i{\kappa_D} \epsilon_{q\, \mu\nu}^B  
\sum_{i=1}^{n}   \left[ 
\frac{k_i^{\nu} q_{\rho}}{k_i \cdot q} ( S_i - {\bar{S}}_i)^{\mu \rho}
  - \frac{1}{2} ( S_i - {\bar{S}}_i )^{\mu \nu} \right]M_n(k_i , \epsilon_i , {\bar{\epsilon}}_i )  + \Ord (q) \, ,
  \label{Orderq0}
}
where 
$\epsilon_{q\, \mu\nu}^B = \frac{1}{2}\left (\epsilon_{q\,\mu} \bar{\epsilon}_{q\, \nu}  -\epsilon_{q\,\nu} \bar{\epsilon}_{q\, \mu} \right )$ is the polarization tensor of the B-field.
This result is consistent with the `holomorphic soft theorem' for the B-field 
found in the bosonic string in Ref. \cite{DMM1} (further details are 
given in Sec. \ref{bosonicsubleading}).

The leading soft theorem for the B-field can be added to the corresponding  expression derived for the dilaton and graviton~\cite{DMM3,DMM1,DMM2} to define a unified operator that collects  the soft behaviour of amplitudes in theories of gravity coupled to a dilaton and a two form. The full soft operator then turns out to be:
\ea{
M_{n+1}= {\kappa_D} \epsilon_{q,\mu}\bar{\epsilon}_{q,\nu}
\sum_{i=1}^n& \Big[\frac{k_i^\mu k_i^\nu}{k_i \cdot q}-\frac{i}{2} \frac{k_i^\mu q_\rho \big(L_i+2\bar{S}_i\big)^{\nu\rho}}{k_i \cdot q}-   
\frac{i}{2} \frac{k_i^\nu q_\rho \big(L_i+2S_i\big)^{\mu\rho}}{k_i \cdot q} 
\nonumber\\
& 
+ \frac{i}{2}  \Big( S^{\mu \nu} -
 {\bar{S}}^{\mu \nu} \Big) \Big]M_n
 + \Ord(q)
 \label{unified}
}
This expression generically reproduces the soft behavior 
of the graviton, dilaton and B-field upon 
symmetrization, respectively, antisymmetrization of the 
polarization vectors $\epsilon_{q,\mu}\bar{\epsilon}_{q,\nu}$. 
As such, it can be considered the soft theorem of 
double-copied Yang-Mills theory.

As we will also show, contrary to the case of the dilaton and graviton, it is not
 possible to fix completely 
the term of order $q$ in the soft behavior of the B-field by this construction.
To explain why, let us note that a caveat in the construction above is that 
the 
 quantity  {$N_{n+1}$} in Eq.~(\ref{decomposition}) may, at a particular order 
in $q$, contain terms local in $q$ 
that are independently gauge invariant. Such terms can for obvious reasons 
not be related to the factorizing set of diagrams by gauge invariance. 
For the graviton and dilaton, this is avoided through 
order $q$ in the soft momentum, 
since the most general local expression for a gauge invariant symmetric 
two-index tensor is of $\Ord(q^2)$~\cite{BDDN}:
{
\ea{
E_S^{\mu \nu} = 
q_\rho q_\sigma
A^{\rho \mu} A^{\sigma \nu}
}
where,  due to gauge invariance, $A^{\rho \mu} = - A^{\mu \rho}$ is an antisymmetric function constructed out of the momenta and polarization vectors of the external states, and is furthermore a local function in $q$.
}

For the antisymmetric B-field, however, things are different, since it is 
possible to write a general expression for an antisymmetric two-index 
tensor local in $q$, which obeys gauge invariance starting already at $\Ord(q)$:
\ea{
E_A^{\mu \nu} = q_{\rho} A^{\rho \mu \nu} \, ,
\label{EA}
}
where $A^{\rho \mu \nu}$ is a totally antisymmetric tensor constructed 
from the external momenta and polarizations, and is furthermore a local 
function in $q$.
For this reason one is not able to constraint {$N_{n+1}$} {through 
order $q$} in the case of a soft B-field. It is nevertheless possible to 
decompose the amplitude through any order in the soft expansion into two separately gauge invariant parts, 
as follows using the notation
 $v_{[\mu} w_{\nu] }= \frac{1}{2} \left( v_{\mu} w_{\nu} - v_{\nu} w_{\mu}
\right)$:
\ea{
M_{n+1} (q, k_i)
= 
&\epsilon_{q\, \mu\nu}^B \,q_\rho\, {A}^{\rho \mu \nu}(q, k_i)
 -i {\kappa_D}\epsilon_{q\, \mu\nu}^B  \sum_{i=1}^{n}
\Bigg \{
\frac{1}{2} \Bigl[
 S_i^{\mu \nu}  -
 \bar{S}_i^{\mu \nu}
 \Bigr]
\\
&
+\frac{1}{k_{i}\cdot q}
\Bigl[
q_{\rho}\Bigl(
k_{i}^{[\mu}\bar{S}_{i}^{\nu]\rho}
+k_{i}^{[\nu}S_{i}^{\mu]\rho}
\Bigr)
+i
q_{\rho} \q_{\sigma}\Bigl(
S_{i}^{\rho [\mu} \bar{S}_{i}^{\nu]\sigma}
\Bigr)
\Bigr]
 \Bigg \}
 M_n (k_i + q , \epsilon_i , {\bar{\epsilon}}_i )  \, .
 \nonumber
}
where one part remains unconstrained due to the preceding discussion, but is local in $q$, 
while  the  other part factorizes as a soft theorem (one can Taylor expand $M_n (k_i +q)$).
The factorizing part encodes the soft theorem, as well as containing all terms needed to gauge covariantize the
first part of Eq.~\eqref{decomposition} involving $M_3$. 
This expression can essentially be seen as the main consequence of the B-field obeying a soft theorem.
As we will show in Sec.~\ref{gauge}, the order $q$ factorizing terms can compactly be written in terms of angular momentum operators.

We will in this work  first derive the above summarized results  and then
explicitly derive the soft behavior of the Kalb-Ramond B-field both in the bosonic string and in superstrings.
We confirm the proposed soft relations, and furthermore we provide explicit 
expression for ${A}^{\rho \mu \nu}$ showing that it is non-zero. 
In particular, we will see that the Bloch-Wigner dilogarithm appears 
in ${A}^{\rho \mu \nu}$ in both string theories, which leads us to
conclude that ${A}^{\rho \mu \nu}$ cannot be written as a soft theorem. 
As regards the field theory limit, the form of ${A}^{\rho \mu \nu}$ remains 
to be understood.

The paper is organized as follows: In Sec.~\ref{gauge} we show that 
on-shell gauge invariance fixes the leading soft behavior {(of order $q^0$)}
 of the B-field in
 tree-level amplitudes, while the subleading part {(of order $q$)}
can only be partially fixed. 
In Sec.~\ref{bosonic} and \ref{super} we explicitly compute amplitudes in 
the bosonic, respectively, the supersymmetric string theory involving 
a soft Kalb-Ramond state to confirm the new soft theorem as well as 
to show that the subleading soft behavior cannot be factorized. 
In Sec.~\ref{conclusion} we provide our conclusions.

\section{Soft theorem for $B_{\mu \nu}$ from gauge invariance}
\label{gauge}

In this section we derive the soft theorem for the antisymmetric tensor 
$B_{\mu \nu}$ in an amplitude with only massless particles, {i.e. Kalb-Ramond fields, gravitons and dilatons.} 
We will see that, unlike for the graviton and dilaton, in the case of the antisymmetric
tensor we can only determine the soft behavior through order $q^0$.
The soft behavior of order $q^1$ cannot be fixed by gauge 
invariance.

Let us start from the pole term given by the diagrams where  the soft particle 
is attached to one of the other external particles, as depicted in Fig.~\ref{factorization}. 
As explained in the introduction their sum is given 
by (we define $M_{n+1} = \epsilon_q^\mu \bar{\epsilon}_q^\nu M_{\mu \nu}$)
\begin{eqnarray}
M_{\mu \nu}^{pole} = {\kappa_D} \sum_{i=1}^n \frac{ [  k_{i\mu} -  i
q^\rho S_{\mu \rho} ][
 k_{i\nu} -  i
q^\sigma {\bar{S}}_{\nu \sigma} ]}{k_i \cdot q} M_n (k_i +q) \, , 
\label{BB1a}
\end{eqnarray}
where $k_i$ and $q$ were put on shell, i.e. $k_i^2 = q^2 = 0$, the polarization tensors $\epsilon_q^\mu \bar{\epsilon}_q^\nu$ were stripped off, and
\ea{
S_{i\, \mu \rho} = i \left(\epsilon_{i \mu}
\frac{\partial}{\partial \epsilon_i^{\rho}} - \epsilon_{i \rho}
\frac{\partial}{\partial \epsilon_i^{\mu}} \right)
~;~{\bar{S}}_{i\, \nu \sigma} = i
\left({\bar{\epsilon}}_{i \nu}
\frac{\partial}{\partial {\bar{\epsilon}}_i^{\sigma}} - {\bar{\epsilon}}_{i \sigma }
\frac{\partial}{\partial {\bar{\epsilon}}_i^{\nu}} \right) \, .
\label{BB1bk}}
In the case of a soft antisymmetric tensor, where $M_{\mu \nu}^{pole}$ is 
antisymmetric under the exchange of the indices $\mu$ and $\nu$, the expression reduces to:
\begin{eqnarray}
M_{\mu \nu}^{pole} = {\kappa_D} \sum_{i=1}^n \frac{ -i   k_{i [\mu} 
q^\sigma {\bar{S}}_{i\,\nu] \sigma} 
-  i  k_{i[ \nu} q^\rho S_{i\, \mu] \rho} 
- q^\rho S_{i\,[\mu \rho}
q^\sigma {\bar{S}}_{ i \,\nu] \sigma} }{k_i \cdot q} M_n (k_i +q) \, , 
\label{BB1b}
\end{eqnarray}
where 
 $v_{[\mu} w_{\nu] }= \frac{1}{2} \left( v_{\mu} w_{\nu} - v_{\nu} w_{\mu}
\right)$.
The previous expression is not gauge invariant, i.e. it is not vanishing
when we saturate it with $q^\mu$ or $q^\nu$.  It is possible, however, to add to it a term, local in $q$,
which will make it gauge invariant, i.e.:
\begin{eqnarray}
 M_{\mu \nu} = {\kappa_D}
\sum_{i=1}^n&& \left[ \frac{ -i   k_{i [\mu} 
q^\sigma {\bar{S}}_{i\,\nu] \sigma} 
-  i  k_{i[ \nu} q^\rho S_{i\, \mu] \rho} 
- q^\rho S_{i\,[\mu \rho}
q^\sigma {\bar{S}}_{ i \,\nu] \sigma} }{k_i \cdot q}  \right. \nonumber \\
&& \left.
+ 
 \frac{i}{2}
 \left( S_{i\,\mu \nu}-  {\bar{S}}_{i \,\mu \nu}\right)  \right]
 M_n (k_i +q) + N_{\mu \nu} (q; k_i) \, ,
\label{alterb} 
\end{eqnarray}
{where $N_{\mu \nu}$ is now the antisymmetric gauge-invariant remainder of the amplitude.}
It is easy to see that the expression in the square bracket above vanishes when
we saturate it with $q^\mu$ or $q^\nu$. Gauge invariance then implies 
the following conditions on the additional term $N_{\mu \nu}$:
\begin{eqnarray}
q^\mu N_{\mu \nu} (q; k_i) = q^\nu N_{\mu \nu} (q; k_i) =0  \, .
\label{gauinvry}
\end{eqnarray}
Expanding around $q=0$, at the lowest order, we get two conditions:
\begin{eqnarray}
q^\mu N_{\mu \nu} (q=0) = q^\nu N_{\mu \nu} (q=0) =0
\label{q=0bis}
\end{eqnarray}
that are {for $ N_{\mu \nu} = -  N_{\nu \mu}$} consistent with 
\begin{eqnarray}
 N_{\mu \nu} (q=0)  = 0 \, .
\label{q=0aab}
\end{eqnarray}
At the next order  in the soft momentum $q$ we get
\begin{eqnarray}
q^\mu q^\rho \frac{\partial}{\partial q^\rho} N_{\mu \nu} (q=0) = 
q^\nu q^\rho  \frac{\partial}{\partial q^\rho} N_{\mu \nu} (q=0) =0 \, ,
\label{qqderivyu} 
\end{eqnarray}
which implies that 
\begin{eqnarray}
\frac{\partial}{\partial q^\rho} N_{\mu \nu} (q=0)= A_{\rho \mu \nu} \, , 
\label{deri94}
\end{eqnarray}
where $A_{\rho \mu \nu}$ 
is a completely antisymmetric tensor under the exchange of the three indices,
and is only a function of the  momenta and polarizations of the hard external 
particles. Notice that the tensor $N_{\mu \nu}$ contains in general higher
powers in the soft momentum $q$.
Since $N_{\mu \nu}(q; k_i) = q^\rho A_{\rho \mu \nu}(k_i) + \Ord(q^2)$ and since we assume 
that it is local in $q$, we may just as well express it as:
\ea{
N_{\mu \nu}(q; k_i) = q^\rho A_{\rho \mu \nu}(q, k_i)  \, ,
\label{Nmunu}
}
to all orders in $q$, and automatically satisfying Eq.~\eqref{gauinvry}, where now $A_{\rho \mu \nu}(q, k_i)$ contains all the higher order terms in $q$.
We end up with:
\begin{eqnarray}
 M_{\mu \nu} = -{\kappa_D}
\sum_{i=1}^n&& \left[ \frac{ i   k_{i [\mu} 
q^\sigma {\bar{S}}_{i\,\nu] \sigma} 
+  i  k_{i[ \nu} q^\rho S_{i\, \mu] \rho} 
+q^\rho S_{i\,[\mu \rho}
q^\sigma {\bar{S}}_{ i \,\nu] \sigma} }{k_i \cdot q}  \right. \nonumber \\
&& \left.
- 
 \frac{i}{2}
 \left( S_{i\,\mu \nu}-  {\bar{S}}_{i \,\mu \nu}\right)  \right]
 M_n (k_i +q) + q^\rho A_{\rho \mu \nu}(q, k_i) \, .
\label{alterb} 
\end{eqnarray}
This is an exact relation between the $n+1$ and $n$-point amplitudes, valid to any order in the soft expansion.
Obviously, since the last term is gauge invariant, ${A}_{\rho \mu \nu}$  cannot be fixed by gauge invariance of the amplitude. 
To conclude, in the case of a soft antisymmetric tensor scattering on other massless states, gauge invariance fixes the amplitude only through the order $q^0$. 
The term of order $q$ contains a totally antisymmetric tensor that cannot be fixed by gauge invariance.  

It is convenient for later use to introduce a new tensor ${\tilde{A}}_{\rho \mu \nu}$
for the leading order expression of $A_{\rho \mu \nu}$, in the following way
\ea{
&A_{\rho \mu \nu}(q=0 ,k_i) = 
{\tilde{A}}_{\rho \mu \nu} (k_i)
\label{substanti}
\\
&
- \frac{i}{2} {\kappa_D} \sum_{i=1}^n 
 \left[ 
\left(S_{i} - {\bar{S}}_{i} \right)_{\mu \nu}
\frac{\partial}{\partial k_i^{\rho} }  - 
\left(S_{i} - {\bar{S}}_{i} \right)_{\rho \nu}
\frac{\partial}{\partial k_i^{\mu} } -
\left(S_{i} - {\bar{S}}_{i} \right)_{\mu \rho}
\frac{\partial}{\partial k_i^{\nu} }\right] M_n (k_i) \nonumber
}
This is possible since the operator in the squared bracket is just another totally antisymmetric tensor. Expanding Eq.~\eqref{alterb} and inserting this alternative expression for $A_{\rho \mu \nu}$ at leading order, we arrive at
\ea{
M_{\mu \nu} = &-i {\kappa_D} \sum_{i=1}^{n}   \Bigg \{ 
\frac{q_{\rho}k_i^{[\nu}( S_i - {\bar{S}}_i)^{\mu] \rho}}{k_i \cdot q} 
  - \frac{1}{2} ( S_i - {\bar{S}}_i )^{\mu \nu} 
   +
q_\rho \Bigl[
 S_i^{\rho [\mu} \partial_i^{\nu]}
 +
 \bar{S}_i^{\rho [\nu} \partial_i^{\mu]}
 \Bigr]
 \nonumber
 \\
&
+\frac{q_{\rho}q_{\sigma}}{k_{i}\cdot q}
\Bigl[
\Bigl(
k_{i}^{[\mu}\bar{S}_{i}^{\nu]\rho}
+k_{i}^{[\nu}S_{i}^{\mu]\rho}
\Bigr)\partial_{i}^{\sigma}
+i
\Bigl(
S_{i}^{\rho [\mu} \bar{S}_{i}^{\nu]\sigma}
\Bigr)
\Bigr]
  \Bigg\}M_n(k_i , \epsilon_i , {\bar{\epsilon}}_i )
  + q_\rho\, \tilde{A}^{\rho \mu \nu}(k_i)
  + \Ord(q^2)
   \label{finalgaugeinvariance}
  }
where $\partial_i^\mu \equiv \partial/\partial k_{i\mu}$.
This expression can be written more compactly, by defining holomorphic and antiholomorphic total angular momentum operator, as follows:
\ea{
J_i^{\mu \nu} = L_i^{\mu \nu} + S_i^{\mu \nu} \, , \quad
\bar{J}_i^{\mu \nu} = L_i^{\mu \nu} + \bar{S}_i^{\mu \nu} \, , \quad
L_i^{\mu \nu} = i \left ( k_i^{\mu }\partial_i^\nu - k_i^\nu \partial_i^\mu \right ) \, ,
}
These operators especially turn useful, when considering the action on superstring amplitude. Let us consider the operator:
\ea{
\epsilon_{q\, \mu\nu}^B \sum_{i=1}^{n}
\frac{q_{\rho}q_{\sigma}}{k_{i}\cdot q}
J_{i}^{\rho [\mu} \bar{J}_{i}^{\nu]\sigma}
= 
\epsilon_{q\, \mu\nu}^B \sum_{i=1}^{n}
\frac{q_{\rho}q_{\sigma}}{k_{i}\cdot q}
\left [
L_{i}^{\rho [\mu} {L}_{i}^{\nu]\sigma}
+
L_{i}^{\rho [\mu} \bar{S}_{i}^{\nu]\sigma}
+
S_{i}^{\rho [\mu} L_{i}^{\nu]\sigma}
+
S_{i}^{\rho [\mu} \bar{S}_{i}^{\nu]\sigma}
\right]
}
Considering the first term on the right-hand side, 
{it can be shown to vanish
due to antisymmetry and transversality of $\epsilon^B_{q \mu \nu}$, as well as the mass-shell condition $q^2=0$.
Therefore we get, after inserting the explicitly expression for $L_i$,}
\ea{
\epsilon_{q\, \mu\nu}^B \sum_{i=1}^{n}
\frac{q_{\rho}q_{\sigma}}{k_{i}\cdot q}
J_{i}^{\rho [\mu} \bar{J}_{i}^{\nu]\sigma}
= &
\epsilon_{q\, \mu\nu}^B \sum_{i=1}^{n}
\Bigg\{
\frac{q_{\rho}q_{\sigma}}{k_{i}\cdot q}
\left [
i(S_{i}^{\rho [\mu} 
-
\bar{S}_{i}^{\rho [\mu})
k_{i}^{\nu]}\partial_i^{\sigma}
+
S_{i}^{\rho [\mu} \bar{S}_{i}^{\nu]\sigma}
\right]
\nonumber \\
& \qquad \qquad 
- i q_\rho (S_{i}^{\rho [\mu} 
-
\bar{S}_{i}^{\rho [\mu})
\partial_i^{\nu]}
\Bigg\}
}
which is exactly equal to the order $q$ factorized part of 
Eq.~\eqref{finalgaugeinvariance}.
In other words, we find a more compact form for the 
expanded amplitude through order $q$, reading: 
\ea{
M_{n+1} = &{\kappa_D} \epsilon_{\mu} \bar{\epsilon}_\nu 
 \sum_{i=1}^{n}  
 \Bigg \{ 
 \frac{i}{2} ( S_i - {\bar{S}}_i )^{\mu \nu} 
+i \frac{q_{\rho}k_i^{[\mu}( S_i - {\bar{S}}_i)^{\nu] \rho}}{k_i \cdot q} 
   +
   \frac{q_{\rho}q_{\sigma}}{k_{i}\cdot q}
J_{i}^{\rho [\mu} \bar{J}_{i}^{\nu]\sigma}
\Bigg \}
M_n(k_i , \epsilon_i , {\bar{\epsilon}}_i )
\nonumber \\
&
+
\epsilon_{\mu} \bar{\epsilon}_\nu \,q_\rho\, \tilde{A}^{\rho \mu \nu}(k_i, \epsilon_i , {\bar{\epsilon}}_i )
+ \Ord (q^2) \, .
\label{compactsoftoperator}
}
where we used that the contraction of $\mu \nu$ is antisymmetric, and we 
can thus equally write $\epsilon_{\mu \nu}^B \to 
\epsilon_\mu \bar{\epsilon}_\nu$. Notice that
we could also trivially rewrite 
the first two terms in the right-hand side of the previous expression 
in terms of $J_i$ and ${\bar{J}}_i$, since $J_i - {\bar{J}}_i = S_i - \bar{S}_i$.
One may wonder whether the part that remains unfixed by gauge 
invariance also factorizes in terms of a soft and a hard part due 
to some other property of the amplitude.
In the subsequent two sections we investigate this question by 
computing explicitly the unfactorized part involving 
$\tilde{A}^{\rho \mu \nu}$ in the bosonic string as well as in 
the superstring. Our conclusion to this question is 
negative, nevertheless, we provide the details and explicit expressions 
that may be useful in other regards. The conclusion about the field theory 
limit of $\tilde{A}^{\rho \mu \nu}$ remains open.

\section{Soft scattering of $B_{\mu \nu}$ in the bosonic string}
\label{bosonic}

For the derivation of the scattering amplitude
involving $n+1$ massless closed string states in the bosonic string we
refer to Sec.~2 in Ref.~\cite{DMM2}. Therein it was shown that the 
$(n+1)$-point amplitude, $M_{n+1}$, can be written as a convolution
\begin{eqnarray}
M_{n+1} =   M_n \ast S \ ,
\label{MMS}
\end{eqnarray}
where $M_{n}$ is just the $n$-point amplitude, and where by $*$ a convolution of integrals is understood, and $S$ carries all the information of the additional external state. The point is that the computation of the soft behavior of $M_{n+1}$ is equivalent to computing the soft expansion of $S$. Let us quote the expressions for $M_n$ and $S$:
\seq{
M_n = &\, \frac{8\pi}{\alpha'}\left (\frac{\kappa_D}{2\pi}\right )^{n-2} 
 \int \frac{\prod_{i=1}^n d^2z_i }{dV_{abc}} \int \left[\prod_{i=1}^n d\f_i \prod_{i=1}^{n} d {\bar{\f}}_i \right]   \prod_{i<j} |z_i - z_j |^{\alpha' k_i k_j}   \\
& \times \exp \left[ -\sum_{i<j}  \frac{\f_i \f_j}{(z_i - z_j)^2}  (\epsilon_i \epsilon_j) + \sqrt{\frac{\alpha'}{2}} \sum_{i \neq j} \frac{ \f_i (\epsilon_i k_j) }{z_i - z_j}  \right]  \\
& \times \exp \left[- \sum_{i<j}  \frac{{\bar{\f}}_i {\bar{\f}}_j}{({\bar{z}}_i - {\bar{z}}_j)^2}  ({\bar{\epsilon}}_i {\bar{\epsilon}}_j) + \sqrt{\frac{\alpha'}{2}} \sum_{i \neq j} \frac{ {\bar{\f}}_i ({\bar{\epsilon}}_i k_j) }{{\bar{z}}_i - {\bar{z}}_j}  \right] ,
\label{nonsoftonly}
}
and
\seq{
S \equiv  
&\, 
\kappa_D
\int \frac{d^2 z}{2 \pi} \,\, \sum_{i=1}^{n} \left(\f_i \frac{ (\epsilon_q \epsilon_i)}{(z-z_i)^2} +    \sqrt{\frac{\alpha'}{2}} \frac{(\epsilon_q k_i)}{z-z_i} \right) \sum_{j=1}^{n} \left({\bar{\f}}_j \frac{ ({\bar{\epsilon}}_q {\bar{\epsilon}}_j)}{({\bar{z}}- {\bar{z}}_j)^2} +    \sqrt{\frac{\alpha'}{2}} \frac{({\bar{\epsilon}}_q k_i)}{{\bar{z}}-{\bar{z}}_i} \right)  \\
& \times \exp \left[ - \sqrt{\frac{\alpha'}{2}}  \sum_{i=1}^{n} \f_i \frac{(\epsilon_i q)  }{z-z_i} \right] \exp \left[ - \sqrt{\frac{\alpha'}{2}}  \sum_{i=1}^{n} {\bar{\f}}_i \frac{({\bar{\epsilon}}_i q)  }{{\bar{z}}-{\bar{z}}_i} \right]\prod_{i=1}^{n} |z- z_i|^{\alpha' q k_i} \, ,
\label{last3lines}
}
where $z_i$ are the Koba-Nielsen variables of the hard states, and $z$ is for the soft state.
Grassmannian variables $\f_i$ have been introduced and in this
notation $\epsilon_i$, the holomorphic polarization vector of the massless
 closed states, are also Grassmannian, and likewise for the antiholomorphic 
counterparts, denoted with a bar. $k_i$ are the momenta of the hard 
states, while $q$ is for the soft state, and $\alpha'$ is the string Regge slope.

The expansion of $S$ was computed in the decomposition:
\begin{eqnarray}
S = \kappa_D \left ( S_1 + S_2 + S_3 \right ) + \Ord(q^2) \ , 
\label{SSi}
\end{eqnarray}
defined by:
\seq{
S_{1} = &\,
\frac{\alpha'}{2} \int \frac{ d^2 z}{2\pi}\, \sum_{i=1}^{n}\frac{(\epsilon_q k_i)}{z-z_i}\sum_{j=1}^{n}  \frac{({\bar{\epsilon}}_q k_j)}{{\bar{z}}-{\bar{z}}_j} \prod_{i=1}^{n} |z- z_i|^{\alpha' q k_i}
\\
& \times 
\Bigg\{ 
1
 - \sqrt{\frac{\alpha'}{2}}
\sum_{k=1}^{n}  \Bigg(
\f_k  \frac{(\epsilon_k q)  }{z-z_k} + {\bar{\f}}_k \frac{({\bar{\epsilon}}_k q)  }{{\bar{z}}-{\bar{z}}_k}
\Bigg) 
+ \frac{1}{2}\left( \frac{\alpha'}{2}\right )
\\
&
\times\Bigg[
\left( \sum_{h=1}^{n}   \f_h  \frac{(\epsilon_h q)  }{z-z_h}  \right)^2
+
\left( \sum_{h=1}^{n}   {\bar{\f}}_h \frac{({\bar{\epsilon}}_h q)  }{{\bar{z}}-{\bar{z}}_h}   \right)^2
+  
2\left( \sum_{h=1}^{n}   \f_h  \frac{(\epsilon_h q)  }{z-z_h}  \right) \left( \sum_{h=1}^{n}   {\bar{\f}}_h \frac{({\bar{\epsilon}}_h q)  }{{\bar{z}}-{\bar{z}}_h}   \right)
\Bigg]
\Bigg\} \, ,
\label{S1}
}
\seq{
S_2 = &\,
\int \frac{ d^2 z}{2\pi} \sum_{i=1}^{n} \left(\f_i \frac{ (\epsilon_q 
\epsilon_i)}{(z-z_i)^2}\right) 
\sum_{j=1}^{n} \left({\bar{\f}}_j \frac{ ({\bar{\epsilon}}_q 
{\bar{\epsilon}}_j)}{({\bar{z}}- {\bar{z}}_j)^2}  \right)\prod_{\ell=1}^{n}
 |z- z_{\ell}|^{\alpha' q k_{\ell}}
 \\
& \times \Bigg\{
1
-  
\sqrt{\frac{\alpha'}{2}}  \sum_{k=1}^{n} \Bigg( 
\f_k \frac{\epsilon_k  q}{z-z_k}
+
{\bar{\f}}_k \frac{({\bar{\epsilon}}_k q)  }{{\bar{z}}-{\bar{z}}_k}
 \Bigg)
 + \frac{1}{2}\left( \frac{\alpha'}{2}\right )
\\
& 
\times \Bigg[
\left( \sum_{h=1}^{n}   \f_h  \frac{(\epsilon_h q)  }{z-z_h}  \right)^2
+
\left( \sum_{h=1}^{n}   {\bar{\f}}_h \frac{({\bar{\epsilon}}_h q)  }{{\bar{z}}-{\bar{z}}_h}   \right)^2
+  
2\left( \sum_{h=1}^{n}   \f_h  \frac{(\epsilon_h q)  }{z-z_h}  \right) \left( \sum_{h=1}^{n}   {\bar{\f}}_h \frac{({\bar{\epsilon}}_h q)  }{{\bar{z}}-{\bar{z}}_h}   \right)
\Bigg]  \Bigg\}\, ,
\label{S2}
}
\seq{
S_3 = &\,
\sqrt{\frac{\alpha'}{2}}\int \frac{ d^2 z}{2\pi} \sum_{i=1}^{n} 
\sum_{j=1}^{n} \left[ \left(  \frac{ \f_i(\epsilon_q \epsilon_i)}{(z-z_i)^2}
\right) \left(\frac{({\bar{\epsilon}}_q k_j)}{{\bar{z}}-{\bar{z}}_j}  \right) +
\left( \frac{ {\bar{\f}}_i({\bar{\epsilon}}_q {\bar{\epsilon}}_i)}{({\bar{z}}- 
{\bar{z}}_i)^2} \right)
\left(\frac{(\epsilon_q k_j)}{z-z_j} \right)  \right] \prod_{\ell=1}^{n} 
|z- z_{\ell}|^{\alpha' q k_{\ell}} 
\\
& \times  \Bigg\{ 
1
-  
\left(  
\frac{\sqrt{2\alpha'}}{2}   \right)   \sum_{k=1}^{n} \Big( 
\f_k \frac{\epsilon_k q}{z-z_k} 
+
{\bar{\f}}_k \frac{({\bar{\epsilon}}_k q)  }{{\bar{z}}-{\bar{z}}_k} 
\Big) 
 + \frac{1}{2}\left( \frac{\alpha'}{2}\right )
\\
& 
\times \Bigg[
\left( \sum_{h=1}^{n}   \f_h  \frac{(\epsilon_h q)  }{z-z_h}  \right)^2
+
\left( \sum_{h=1}^{n}   {\bar{\f}}_h \frac{({\bar{\epsilon}}_h q)
  }{{\bar{z}}-{\bar{z}}_h}   \right)^2
+  
2\left( \sum_{h=1}^{n}   \f_h  \frac{(\epsilon_h q)  }{z-z_h}  \right) \left( \sum_{h=1}^{n}   {\bar{\f}}_h \frac{({\bar{\epsilon}}_h q)  }{{\bar{z}}-{\bar{z}}_h}   \right)
\Bigg]  \Bigg\}\, .
  \label{S_3}
  }
Each part was further split in $S_i^{(a)}$, a=0,1,2, with the 
index $a$  labelling  the order of expansion in $q$ of the integrand modulo 
the factor $|z-z_l|^{\aqk{l}}$, which has to be integrated. The explicit results 
for $S_i^{(a)}$  can be found in Ref.~\cite{DMM2}, and {apply 
to any massless closed-string state}. Here we are interested in the antisymmetric part in $\epsilon_q^\mu \bar{\epsilon}_q^\nu$ of those expressions.
{First, the result for $S_1^{(0)}$ is:}
\ea{
S_1^{(0)}|_B =&\, 
 \epsilon_{q}^{B\mu\nu} \sum_{i\neq j\neq m}^n k_{i\mu}k_{j\nu}
\left (\frac{\alpha'}{2}\right )^2  (q k_m) 
\Bigg [{\rm Li}_2\left( \frac{\bar{z}_i-\bar{z}_m}{\bar{z}_i-\bar{z}_j}\right)-{\rm Li}_2\left(\frac{z_i-z_m}{z_i-z_j }\right)
\nonumber\\
& 
+
\log\frac{|z_i-z_j|}{|z_i-z_m|}\log\left(\frac{z_m-z_j}{\bar{z}_m-\bar{z}_j}
\frac{\bar{z}_i-\bar{z}_j}{z_i-z_j}\right) \Bigg ]
+ \Ord(q^2)
\ .
\label{S10B}
}
We can show that the part in the square bracket is just the Bloch-Wigner Dilog, which is analytic and continuous%
\footnote{We thank Lance Dixon for pointing this out to us during the Nordita program Aspects of Amplitudes.}.
 Denoting $\zeta \equiv (\zz{i}{m})/(\zz{i}{j})$ we can write the square bracket of the antisymmetric part as:
\ea{
 {\rm Li}_2(\bar{\zeta}) - {\rm Li}_2 (\zeta) - \log |\zeta| \log\left (\frac{1-\zeta}{1-\bar{\zeta}} \right )
& = - 2 i({\rm Im} ( {\rm Li}_2(\zeta) ) + {\rm arg}(1-\zeta) \log |\zeta | ) \nonumber \\
& = - 2i D_2 (\zeta)
}
where in the second line we identified the Bloch-Wigner dilog, denoted as $D_2$.
This function has the following properties (as well as many other, not relevant here):
\begin{itemize}
\item It is a real function on $\mathbb{C}$, and analytic except at the points $\zeta = \{0,1\}$, where it is only continuous, but not differentiable. For us, $\zeta$ never takes these values, since $z_i \neq z_m \neq z_j$.

\item It has a six-fold symmetry: 

$D_2 (\zeta) = D_2(1- \zeta^{-1}) = D_2 \left (\frac{1}{1-\zeta} \right ) = - D_2 (\zeta^{-1}) = - D_2 (1-\zeta) = - D_2 \left ( \frac{\zeta}{\zeta-1}\right ) 
$

and furthermore $ D_2 (\bar{\zeta}) = -D_2 (\zeta)$, thus $D_2(\mathbb{R}\backslash \{0,1\} ) = 0$.
\end{itemize}
The antisymmetric part of $S_1^{(0)}$ can thus be written as
\ea{
S_1^{(0)}|_B ={i}
 \epsilon_{q}^{B\mu\nu} \sum_{i\neq j\neq m}^n (k_{i \nu}k_{j\mu}-k_{i\mu}k_{j\nu})
\left (\frac{\alpha'}{2}\right )^2  (q k_m)  D_2\left (\frac{\zz{i}{m}}{\zz{i}{j}} \right ) 
}
 It can be checked that this expression is gauge-invariant by itself by using e.g. the relation $D_2 (\zeta) = D_2 (\frac{1}{1-\zeta}) $. In fact, we can write it in the form of Eq.~\eqref{EA}, making it explicitly gauge invariant:
\ea{
S_1^{(0)}|_B =
 i\epsilon_{q\mu\nu}^{B}  \left (\frac{\alpha'}{2}\right )^2 q_\rho \sum_{i\neq j\neq m}^n \frac{2}{3} 
\left [ k_{m}^{\rho} k_{i}^{ [\nu}k_{j}^{\mu]} + k_{m}^{\nu} k_{i}^{ [\mu}k_{j}^{\rho]} 
+ k_{m}^{\mu} k_{i}^{ [\rho}k_{j}^{\nu]}
\right ]
 D_2\left (\frac{\zz{i}{m}}{\zz{i}{j}} \right ) \, ,
}
where the symmetry properties of $D_2$ were used to completely antisymmetrize the expression in the square bracket in the indices $\rho \nu \mu$.

It is convenient also for later use to introduce the tensorial function:
\ea{
T^{\rho \mu \nu}(V,X,Y) = 
\frac{1}{2}&\left (
V^\rho X^\mu Y^\nu-V^\rho X^\nu Y^\mu + V^\mu X^\nu Y^\rho 
\right.
\nonumber \\
&\left.
- V^\mu X^\rho Y^\nu - V^\nu X^\mu Y^\rho + V^\nu X^\rho Y^\mu
\right )
\label{Atensor}
}
which is totally antisymmetric in its indices $\rho \mu \nu$ and in its variables $V, X, Y$.

In terms of this function, we compactly have:
\ea{
S_1^{(0)}|_B^{\mu \nu} =
\frac{2i}{3} \left (\frac{\alpha'}{2}\right )^2\sum_{i\neq j\neq m}^n 
q_\rho  T^{\rho \mu \nu}(k_i, k_j, k_m)
 D_2\left (\frac{\zz{i}{m}}{\zz{i}{j}} \right ) \, .
}
where we stripped off the polarization tensor.

{The antisymmetric part of all other $S_i^{(a)}$ is simply obtained by antisymmetrizing the expressions derived in Ref.~\cite{DMM1, DMM2} in the polarization indices of the soft state, leading to:}
\ea{
S_1^{(1)} |_B^{\mu \nu} = &
 \sqrt{\frac{\alpha'}{2}}
 \sum_{i \neq j}\Bigg\{
  \Ered{
  \frac{k_i^{[\mu} k_j^{\nu]}}{\qk{i}}
 \frac{\teq{i}}{\zz{i}{j}}
 \Bigg [  1+
\sum_{l\neq i} \aqk{l} \log|\zz{i}{l}| 
\Bigg]
}
\nonumber \\
&
+
 \alpha' k_i^{[\mu} k_j^{\nu]}\Bigg [
\frac{ \teq{i}}{2(\zz{i}{j})}
+ 
\sum_{l\neq i} \frac{\teq{l}}{\zz{i}{l}}\log|\zz{i}{j}|
-\sum_{l\neq i,j} \frac{\teq{l} }{\zz{i}{l}} \log|\zz{j}{l}|
\Bigg]
\Bigg \}
+ \text{c.c.}
\label{S11B}
\\
S_1^{(2)} |_B^{\mu \nu}= &
- \frac{\a'}{2} \sum_{i\neq j}^n\left (
\frac{k_{i}^{[\mu} k_j^{\nu]}}{\qk{i}}
\frac{  (\f_i \epsilon_i q)}{  (\zz{i}{j})} \sum_{l \neq i} 
\left ( \frac{\teq{l}}{(z_i-z_l)} + \frac{\tbeq{l}}{(\zbzb{i}{l})}\right )
  + \text{c.c.} \right ) 
  \nonumber \\
  &
  + \frac{\a'}{2} \sum_{i\neq j\neq l}^n
    \frac{k_{j}^{[\mu} k_l^{\nu]}}{\qk{i}}
  \frac{(\teq{i})(\tbeq{i})}{(\zz{i}{j})(\zbzb{i}{l})}
 \label{S12B}
 }
\seal{S2B}{
S_2^{(0)} |_B^{\mu \nu}=&\,
\frac{\alpha'}{2}
 \sum_{i \neq j}^n 
\sum_{l\neq i}\Bigg \{
\Ered{
 \frac{(\qk{j})(\qk{l})}{\qk{i}}
\frac{(\f_i \epsilon_i^{[\mu})(\bar{\f}_i \bar{\epsilon}_i^{\nu]})}{(\zz{i}{j})(\zbzb{i}{l})}
}
\nonumber \\
&
+
 \frac{\te{[\mu}{i}\tbe{\nu]}{j} (\qk{l})+\te{[\mu}{l}\tbe{\nu]}{i} (\qk{j}) +\te{[\mu}{l}\tbe{\nu]}{j} (\qk{i})}{(\zz{i}{l})(\zbzb{i}{j})}
\Bigg\} \, , 
\label{S20B}
\\[5mm]
S_2^{(1)}|_B^{\mu \nu} =&\,
\sqrt{\frac{\a'}{2}}\sum_{i\neq j}^n 
\sum_{l\neq i}
\Bigg\{
\Ered{
 \frac{\qk{l}}{\qk{i}} \frac{
\left (\te{[\mu}{i}  (\teq{j})
-
\te{[\mu}{j}  (\teq{i}) \right )\tbe{\nu]}{i} 
}{(\zz{i}{j})^2(\zbzb{i}{l})}
}
\nonumber \\
&\,
 +\frac{
\left (\te{[\mu}{i}  (\teq{l})
-
\te{[\mu}{l}  (\teq{i}) \right )\tbe{\nu]}{j} 
}{(\zbzb{i}{j}) (\zz{i}{j})^2} 
\Bigg \}
+ \text{c.c.} \, ,
\label{S21B}
\\[5mm]
S_2^{(2)}|_B^{\mu \nu} =&\,
\Ered{\sum_{i \neq j }\frac{1}{\qk{i}}
\sum_{l\neq i}
\left (\frac{(\f_i \epsilon_i^{[\mu})(\f_j \epsilon_j q)- (\f_j \epsilon_j^{[\mu})(\f_i \epsilon_i q)}{(\zz{i}{j})^2}\right )
\left (
\frac{ (\bar{\f}_i \bar{\epsilon}_i^{\nu]})
(\bar{\f}_l \bar{\epsilon}_l q)- (\bar{\f}_l\bar{\epsilon}_l^{\nu]})
(\bar{\f}_i\bar{\epsilon}_i q)}
{ (\zbzb{i}{l})^2}
\right )
}  \, ,
\label{S22B}
}
\seal{S3B}{
S_3^{(0)}|_B^{\mu \nu} = &\,
\sqrt{\frac{\alpha'}{2}}
 \sum_{i\neq j}^n 
 \Bigg [
 \Ered{\frac{ k_i^{[\nu} \te{\mu]}{i}}{\zz{i}{j}} \frac{\qk{j}}{\qk{i}}} 
  - \frac{k_j^{[\nu} \te{\mu]}{i}}{\zz{i}{j}}
+\frac{\a'}{2}  \frac{ k_i^{[\nu} \te{\mu]}{i} \qk{j} - k_j^{[\nu} \te{\mu]}{i} \qk{i}}{\zz{i}{j}}
\nonumber \\
&
+\a'\sum_{l\neq i} 
\frac{[(\qk{l}) k_i^{[\nu}  -(\qk{i}) k_l^{[\nu} ]\te{\mu]}{j}-(\qk{j})k_l^{[\nu}\te{\mu]}{i}}{\zz{i}{j}}
\LN{i}{l}
\nonumber \\
&
+\Ered{ \a'\sum_{l\neq i} 
\frac{(\qk{j})(\qk{l})}{\qk{i}} 
\frac{k_i^{[\nu}\te{\mu]}{i}}{\zz{i}{j}}
\LN{i}{l}
}
\Bigg ]+\text{c.c.} \, ,
 \label{S30B}
\\[5mm]
S_3^{(1)} |_B^{\mu \nu} = &\,
 \sum_{i\neq j}^n  \Bigg\{
 \Ered{
\frac{ (\f_j \epsilon_j q)\te{[\mu}{i}}{(z_i - z_j)^2}
\left ( \frac{ k_{i}^{\nu]}}{k_i q} - \frac{ k_{j}^{\nu]}}{k_j q} \right )}
\Ered{  -\a' 
  \frac{ (\f_i \epsilon_i q)\te{[\mu}{j}k_i^{\nu]}}{(z_i - z_j)^2}\sum_{l\neq i} 
 \frac{\qk{l}}{\qk{i}}\LN{i}{l}
 }
  \nonumber \\
 &
 \Ered{ 
 -\a' 
  \sum_{l\neq i}
 \frac{\qk{l}}{\qk{i}} 
\frac{ (\f_j \epsilon_j q)\te{[\mu}{i}k_i^{\nu]}}{\zz{i}{j}}
 \left (
\frac{1}{2(\zz{i}{l})} - 
\frac{\LN{i}{l}}{(z_i - z_j)}  
 \right) 
 }
 \nonumber \\
 &
 \Ered{
  - \frac{\alpha'}{2}\sum_{l\neq i}
 \frac{\qk{l}}{\qk{i}} 
\frac{\te{[\mu}{i}(k_i^{\nu]} \bar{\f}_j \bar{\epsilon}_j q
+  k_j^{\nu]} \bar{\f}_i \bar{\epsilon}_i q)}
{(\zbzb{i}{j})(\zz{i}{l})}
}
  \nonumber \\
 &
 +\a' \sum_{l\neq i}\
  \frac{ (\f_i \epsilon_i q)\te{[\mu}{j} k_l^{\nu]}}{(z_i - z_j)^2} \LN{i}{l}
  \nonumber \\
 &
+\a' 
 \sum_{l\neq i}
\frac{ (\f_j \epsilon_j q) \te{[\mu}{i}k_l^{\nu]}}{\zz{i}{j}}
 \left (
\frac{1}{2(\zz{i}{l})} - 
\frac{\LN{i}{l}}{(z_i - z_j)}  
 \right) 
 \nonumber \\
 &
 -\frac{\a'}{2} \sum_{l\neq i} 
\frac{\te{[\mu}{j}({k_l^{\nu]} \tbeq{i}} + {k_i^{\nu]} \tbeq{l})}}{(\zz{i}{j})(\zbzb{i}{l})}
 \Bigg\} +\text{c.c.} \, ,
  \label{S31B}
 \\[5mm]
 S_3^{(2)}|_B^{\mu \nu} = &\,
 \Ered{
\sqrt{\frac{\alpha'}{2}} \sum_{i\neq j}^n 
\frac{1}{\qk{i}}
\Bigg [ \sum_{l\neq i,j}^n 
\frac{(\f_l \epsilon_l q) ( \te{[\mu}{j}k_i^{\nu]}
(\f_i \epsilon_i q) 
- \te{[\mu}{i}k_i^{\nu]}(\f_j \epsilon_j q) )
  }{(\zz{i}{j})^2(\zz{i}{l})} 
}
  \nonumber \\
&
\Ered{
+
\frac{ \te{[\mu}{j}(\teq{i}) -\te{[\mu}{i}(\teq{j})}{(\zz{i}{j})^2} 
\sum_{l\neq i} 
\frac{\left (k_i^{\nu]} \tbeq{l} + k_l^{\nu]} \tbeq{i} \right )}{ (\zbzb{i}{l})} 
 \Bigg ] + {\rm c.c.} \, .
}
\label{S32B}
}
{We note that when taking the complex conjugate one must also exchange the indices $\mu \leftrightarrow \nu$, since $\overline{\epsilon_{q\mu \nu}^B} =  - \epsilon_{q\mu \nu}^B = \epsilon_{q\nu \mu}^B$, which follows from the decomposition $\epsilon_{q\,\mu\nu}=\epsilon_{q\mu}\bar{\epsilon}_{q\,\nu}$. }

\subsection{The soft theorem}
\label{bosonicsubleading}
The terms of $\Ord(q^0)$ appear in Eq.~\eqref{S11B}, \eqref{S30B} and 
\eqref{S31B} only.
Summarizing, they read:
\ea{
S|_B^{\mu \nu} =&
 \sum_{i \neq j}\Bigg\{
  \sqrt{\frac{\alpha'}{2}}
  \frac{k_i^{[\mu} k_j^{\nu]}}{\qk{i}}
 \frac{\teq{i}}{\zz{i}{j}}
+
 \sqrt{\frac{\alpha'}{2}}
\frac{ k_i^{[\nu} \te{\mu]}{i}}{\zz{i}{j}} \frac{\qk{j}}{\qk{i}}
  - 
   \sqrt{\frac{\alpha'}{2}}
   \frac{k_j^{[\nu} \te{\mu]}{i}}{\zz{i}{j}}
   \nonumber \\
   &
  +
  \frac{ ((\f_j \epsilon_j q)\te{[\mu}{i} - (\f_i \epsilon_i q)\te{[\mu}{j})k_{i}^{\nu]}}{(k_i q)(z_i - z_j)^2}
\Bigg\} + \text{c.c} + \Ord(q)
\label{SBmnq0}
}
The soft theorem proposed to reproduce this is:
\ea{
M_{B}^{\mu \nu} = -i  \sum_{i=1}^{n}   \left[ 
\frac{k_i^{[\nu} q_{\rho}   ( S_i - {\bar{S}}_i)^{\mu] \rho} }{q k_i}
  - \frac{1}{2} ( S_i - {\bar{S}}_i )^{\mu \nu} \right]M_n(k_i , \epsilon_i , {\bar{\epsilon}}_i )  + \Ord (q).
  \label{MBmnq0}
 }
Using that
\ea{
-i S_i^{\mu \rho} M_n = 
M_n \ast \left [\sum_{j\neq i} \frac{\te{\mu}{i} \te{\rho}{j} - \te{\mu}{j} \te{\rho}{i}}{(\zz{i}{j})^2}
+
\sqrt{\frac{\alpha'}{2}}
\sum_{j\neq i} \frac{\te{\mu}{i} k_j^\nu - k_j^\mu \te{\rho}{i}}{\zz{i}{j}}
\right ]
}
it is straightforward to see that Eq.~\eqref{MBmnq0} exactly reproduces Eq.~\eqref{SBmnq0}.
For checking this, we note that the terms with two $\f$'s  produced
 by $( S_i - {\bar{S}}_i )^{\mu \nu} $ vanishes over the sum, due to opposite 
parity of the numerator and denominator in the exchange of $i\leftrightarrow j$.

We additionally like to make the remark that the soft operator in Eq.~\eqref{MBmnq0} follows also from earlier considerations in Ref.~\cite{DMM3,DMM1}, where it was noticed
that the explicit result in Eq.~\eqref{SBmnq0} can be reproduced by the following {`holomorphic' soft theorem}: By separating the string amplitude into a holomorphic and an antiholomorphic part, and promoting the momentum in the antiholomorphic sector to a (spurious) `antiholomorphic' momentum, $k \to \bar{k}$, it can be shown that both the bosonic string amplitude and the superstring amplitude at $\Ord(q^0)$ can, for any soft state, be equivalently written as:
\ea{
M_{n+1}&=-i \epsilon_\mu \bar{\epsilon}_\nu \sum_{i=1}^n
\left[\frac{q_\rho \bar{k}_i^\nu(L_i+S_i)^{\mu\rho}}{qk_i}
+\frac{q_\rho k_i^\mu(\bar{L}_i+\bar{S}_i)^{\nu\rho}}{qk_i}\right]M_n(k_i,
\epsilon_i;\bar{k}_i,\bar{\epsilon}_i)\Big|_{k=\bar{k}}+{\cal O}(q)
}
where $\bar{L}_i$ denotes the antiholomorphic angular momentum operator that acts on the $\bar{k}$ quantities. The notation $|_{k=\bar{k}}$ means that the $\bar{k}$ is, after the action, identified again with the physical momentum $k$. In Ref.~\cite{DMM3} it was shown that when the soft state is symmetrically polarized, i.e. $ \epsilon_\mu \bar{\epsilon}_\nu \to \epsilon_{\mu \nu}^S$, then the above expression is easily seen to match the known subleading graviton soft theorem, while in Ref.~\cite{DMM1} it was remarked that for an antisymmetric soft state the expression reduces to:
\ea{
M_{n+1}&=-i \epsilon_\mu \bar{\epsilon}_\nu \sum_{i=1}^n\left[ \frac{1}{2} (L_i-\bar{L}_i)^{\mu\nu} +\frac{k_i^\nu q_\rho}{k_iq} (S_i-\bar{S}_i)^{\mu\nu} \right]M_n(k_i,
\epsilon_i;\bar{k}_i,\bar{\epsilon}_i)\Big|_{k=\bar{k}}+{\cal O}(q)
\label{generalorderq0operator}
}
Now, notice that under a gauge transformation for the Kalb-Ramond field,   
$\epsilon_{q \, \mu \nu}^{B}  \rightarrow   \epsilon_{q \, \mu \nu}^{B} + q_{\mu} \chi_{\nu} - q_{\nu} \chi_{\mu}$, the amplitude changes as follows
\begin{eqnarray}
M_{n+1}  \rightarrow  M_{n+1}  
+iq_\rho\chi_\mu \sum_{i=1}^n \Big[(L_i+ S_i)^{\mu\rho}- (\bar{L}_i+\bar{S}_i)^{\mu\rho}\Big]
   M_n ( k_i , \epsilon_i ; {\bar{k}}_i , {\bar{\epsilon}}_i )\Big |_{k=\bar{k}} \, .
\end{eqnarray}
Since for any $q_\rho$ and any $\chi_\mu$ the additional term has to vanish, the following identity must hold
\begin{eqnarray}
\Big. \sum_{i=1}^n (L_i- \bar{L}_i)^{\mu\rho}M_n( k_i , \epsilon_i ; {\bar{k}}_i , {\bar{\epsilon}}_i )\Big |_{k=\bar{k}}=\Big. \sum_{i=1}^n (\bar{S}_i- {S}_i)^{\mu\rho}M_n( k_i , \epsilon_i ; {\bar{k}}_i , {\bar{\epsilon}}_i )\Big |_{k=\bar{k}} \, ,
\end{eqnarray}
which can be checked by a direct calculation. From this we conclude 
that Eq.~\eqref{generalorderq0operator} for an antisymmetric soft state 
reduces to:
\ea{
M_{n+1}=-i \epsilon_{\mu \nu}^B \sum_{i=1}^n
\left[
\frac{1}{2} ({S}_i^{\nu \mu} - \bar{S}_i^{\nu \mu} )
+\frac{q_\rho k_i^{[\nu}(S_i-\bar{S}_i)^{\mu]\rho}}{qk_i}\right]M_n(k_i,
\epsilon_i;\bar{k}_i\bar{\epsilon}_i)\Big|_{k=\bar{k}}+{\cal O}(q)
}
Since this expression no longer involves the $\bar{L}$ operator, we may readily identify $\bar{k} = k$, making $M_n(k_i,\epsilon_i;\bar{k}_i\bar{\epsilon}_i) = M_n$ the physical $n$-point amplitude. This reproduces, and thus confirms once again, the soft theorem in Eq.~\eqref{MBmnq0}.

\subsection{The order $q$ soft behavior}
While there does not exist a complete soft operator reproducing the terms at $\Ord(q)$,
we are still able to greatly reduce the terms into a gauge invariant part that factorizes, and a gauge invariant part that can be written in terms of a totally antisymmetric tensor.

The soft behavior is proposed to admit the form:
\ea{
M^{\mu \nu} = &-i \sum_{i=1}^{n}   \Bigg \{ 
\frac{q_{\rho}q_{\sigma}}{k_{i}\cdot q}
\Bigl[
\Bigl(
k_{i}^{[\mu}\bar{S}_{i}^{\nu]\rho}
+k_{i}^{[\nu}S_{i}^{\mu]\rho}
\Bigr)\partial_{i}^{\sigma}
+i
\Bigl(
S_{i}^{\rho [\mu} \bar{S}_{i}^{\nu]\sigma}
\Bigr)
\Bigr]
 \\
&
+
q_\rho \Bigl[
 S_i^{\rho [\mu} \partial_i^{\nu]}
 +
 \bar{S}_i^{\rho [\nu} \partial_i^{\mu]}
 \Bigr]
  \Bigg\}M_n(k_i , \epsilon_i , {\bar{\epsilon}}_i )
  + q_\rho\, \tilde{A}^{\rho \mu \nu}(k_i)
  + \Ord(q^2) \nonumber
  }
  
  The action of the soft operators on the lower-point amplitude read:
\ea{
&- i \sum_{i=1}^n \frac{q_\rho q_\sigma}{k_i\cdot q} \Bigl(
k_{i}^{[\mu}\bar{S}_{i}^{\nu]\rho}
+k_{i}^{[\nu}S_{i}^{\mu]\rho}
\Bigr)\partial_{i}^{\sigma} M_n
=
M_n \ast
 \sum_{i\neq j} \frac{k_i^{[\nu}}{k_i\cdot q} 
 \nonumber \\
 &\times
 \left [ \aqk{j} \log|\zz{i}{j}| - \sqrt{\frac{\alpha'}{2}} \frac{\teq{j}}{\zz{i}{j}} -\sqrt{\frac{\alpha'}{2}} \frac{\tbeq{j}}{\zbzb{i}{j}} \right ]
\nonumber \\
&\times
\sum_{l\neq i} \left[ 
\frac{\te{\mu]}{i} (\teq{l})-(\teq{i})\te{\mu]}{l}}{(\zz{i}{l})^2} + \sqrt{\frac{\alpha'}{2}} \frac{\te{\mu]}{i} (k_l\cdot q) - (\teq{i})k_l^{\mu]}}{\zz{i}{l}}
\right ] + \text{c.c}
\label{SMn}
\\[5mm]
&\sum_{i=1}^n \frac{q_\rho q_\sigma}{k_i\cdot q} S_i^{\rho[\mu} \bar{S}_i^{\nu] \sigma} M_n
=M_n \ast
 \sum_{i=1} \frac{1}{k_i\cdot q} 
 \nonumber \\
 &\times
\sum_{j\neq i} \left [ 
 \frac{\te{[\mu}{i}(\teq{j})-(\teq{i})\te{[\mu}{j}}{(\zz{i}{j})^2} 
 + \sqrt{\frac{\alpha'}{2}} \frac{\te{[\mu}{i} (k_j\cdot q) - (\teq{i})k_j^{[\mu}}{\zz{i}{j}}
 \right ]
\nonumber \\
&\times
\sum_{l\neq i} \left[ 
\frac{\tbe{\nu]}{i} (\tbeq{l})-(\tbeq{i})\tbe{\nu]}{l}}{(\zbzb{i}{l})^2} + \sqrt{\frac{\alpha'}{2}} \frac{\tbe{\nu]}{i} (k_l\cdot q) - (\tbeq{i})k_l^{\nu]}}{\zbzb{i}{l}}
\right ]
\label{SSbarMn}
\\[5mm]
&-i \sum_{i=1}^n q_\rho q_\rho \Bigl[
 S_i^{\rho [\mu} \partial_i^{\nu]}
 +
 \bar{S}_i^{\rho [\nu} \partial_i^{\mu]}
 \Bigr]
M_n = M_n \ast
 \sum_{i\neq j} \left [ \a' k_j^{[\nu} \log|\zz{i}{j}| - \sqrt{\frac{\alpha'}{2}} \frac{\te{[\nu}{j}}{\zz{i}{j}} -\sqrt{\frac{\alpha'}{2}} \frac{\tbe{[\nu}{j}}{\zbzb{i}{j}} \right ]
\nonumber \\
&\times
\sum_{l\neq i} \left[ 
\frac{(\teq{i})\te{\mu]}{l}-\te{\mu]}{i} (\teq{l})}{(\zz{i}{l})^2} + \sqrt{\frac{\alpha'}{2}} \frac{ (\teq{i})k_l^{\mu]}-\te{\mu]}{i} (k_l\cdot q)}{\zz{i}{l}}
\right ]+ \text{c.c.}
\label{SdMn}
}
It is now a straightforward, but very tedious task, to show that 
all terms in Eq.~\eqref{S11B}-\eqref{S32B}, which have a $1/q$-pole 
exactly matches the terms given by Eq.~\eqref{SMn} and \eqref{SSbarMn}.
These simply come from the Feynman diagrams where the soft state is attached to an external leg. This confirms that all other Feynman diagrams only produce terms that are local in $q$.

Most of the terms given by Eq.~\eqref{SdMn} also matches similar terms in 
the explicit expressions Eq.~\eqref{S11B}-\eqref{S32B}.
However, the term involving  $\f_i \f_j \f_l$ in Eq.~\eqref{SdMn}
 does not have a counterpart
 in the explicit expressions. It therefore has to be gauge invariant on its own, and indeed:
\ea{
-i \sum_{i=1}^n q_\rho S_i^{\rho [\mu} \partial_i^{\nu]}M_n |_{\f\f\f}
=& \sqrt{\frac{\a'}{2}} \sum_{i\neq j} \sum_{l\neq i}
\frac{\te{[\nu}{j}(\te{\mu]}{i} (\teq{l}) - \te{\mu]}{l}\teq{i})}{(\zz{i}{j})(\zz{i}{l})^2}
\nonumber \\
=&
  \sqrt{\frac{\a'}{2}} \sum_{i\neq j} \sum_{l\neq i, j}
 \frac{\te{[\nu}{j}\te{\mu]}{i} (\teq{l}) }{(\zz{i}{j})(\zz{j}{l})(\zz{i}{l})}
 \nonumber \\
 =&
 \frac{1}{3} \sqrt{\frac{\a'}{2}} \sum_{i\neq j} \sum_{l\neq i, j}
\frac{
q_\rho T^{\rho \mu \nu}(\f_i \epsilon_i, \f_j \epsilon_j ,{\f}_l {\epsilon}_l)
}{(\zz{i}{j})(\zz{j}{l})(\zz{l}{i})} \, , 
 \label{ttt}
}
To arrive at the second line we used that the first line vanishes for $l=j$, because $\f_j \f_j = 0$, and for the third line we used the definition in Eq.~\eqref{Atensor}, making the expression explicitly gauge invariant.

The terms with two $\f$'s coming from Eq.~\eqref{SdMn} also do not match to the explicit expression, however, as we will see below, they ensure gauge invariance of what remains, when we subtract the terms coming from Eq.~\eqref{SMn}-\eqref{SdMn} from the explicit expressions. Indeed, our most reduced expression for the soft behavior through the order $q$ reads:
\ea{
M_n &\ast (S_1 + S_2 + S_3)|_B^{\mu \nu}  
 \nonumber \\
 =&
-i\sum_{i=1}^{n}
\left \{
\frac{k_i^{[\nu} q_{\rho}}{q k_i } ( S_i - {\bar{S}}_i)^{\mu] \rho}
  - \frac{1}{2} ( S_i - {\bar{S}}_i )^{\mu \nu}
  \right. \nonumber \\
  & \left.
+\frac{q_{\rho}q_{\sigma}}{k_{i}\cdot q}
\Bigl[
\Bigl(
k_{i}^{[\mu}\bar{S}_{i}^{\nu]\rho}
+k_{i}^{[\nu}S_{i}^{\mu]\rho}
\Bigr)\partial_{i}^{\sigma}
+i
\Bigl(
S_{i}^{\rho [\mu} \bar{S}_{i}^{\nu]\sigma}
\Bigr)
\Bigr]
+
 q_\rho \Bigl[
 S_i^{\rho [\mu} \partial_i^{\nu]}
 +
 \bar{S}_i^{\rho [\nu} \partial_i^{\mu]}
 \Bigr]
 \right \}
 M_n
 \nonumber \\
 &
+
M_n\ast \Bigg[
i \left (\frac{\alpha'}{2}\right )^2\sum_{i\neq j\neq m}^n 
2 k_i^{[\nu}k_j^{\mu]}
  (q k_m)  D_2\left (\frac{\zz{i}{m}}{\zz{i}{j}} \right ) 
  \nonumber \\
  &
  +
\left (\frac{\a'}{2}\right )^{3/2} \sum_{i\neq j}
\frac{k_i^{[\mu}k_j^{\nu]} (\teq{i})
+k_i^{[\nu} \te{\mu]}{i} (\qk{j}) - k_j^{[\nu} \te{\mu]}{i} (\qk{i})
}{\zz{i}{j}} + \text{c.c.}
\nonumber \\
&+
\left (\frac{\a'}{2}\right )^{3/2} \sum_{i\neq j} \sum_{l \neq i}
\frac{ k_i^{[\mu}k_l^{\nu]} \teq{j} + k_i^{[\nu}\te{\mu]}{j} (\qk{l}) - k_l^{[\nu} \te{\mu]}{j} (\qk{i})}{\zz{i}{j}} \LN{i}{l}^2+ \text{c.c.}
\nonumber \\
&+
\frac{\a'}{2} \sum_{i\neq j} \sum_{l \neq i}
\frac{\te{[\mu}{j} \tbe{\nu]}{l}(\qk{i}) - \te{[\mu}{j}k_i^{\nu]} (\tbeq{l}) - \tbe{[\nu}{l} k_i^{\mu]} \teq{j}}{(\zz{i}{j})(\zbzb{i}{l})}
\nonumber \\
&+
\frac{\a'}{2} \sum_{i\neq j} \sum_{l\neq i}
 \frac{(\teq{j}) \te{[\mu}{i}k_l^{\nu]} + (\teq{i}) \te{[\nu}{j} k_l^{\mu]} - 
\te{[\nu}{j}\te{\mu]}{i} (\qk{l})}{(\zz{i}{j})(\zz{i}{l})} + \text{c.c.}
 \nonumber \\
&-
\sqrt{\frac{\a'}{2}} \sum_{i\neq j} \sum_{l\neq i}
\frac{\te{[\nu}{j}(\te{\mu]}{i} (\teq{l}) - \te{\mu]}{l}(\teq{i}))}{(\zz{i}{j})(\zz{i}{l})^2}+ \text{c.c.}
\Bigg]
\label{rest}
 } 
We can express this more compactly in terms of the totally antisymmetric tensor given in Eq.~\eqref{Atensor}:
\ea{
M_n &\ast (S_1 + S_2 + S_3)|_B^{\mu \nu}  
 \nonumber \\
 =&
\,  q_\rho \tilde{A}_{\rm bosonic}^{\rho \mu \nu}
-i\sum_{i=1}^{n}
\left \{
\frac{k_i^{[\nu} q_{\rho}}{q k_i } ( S_i - {\bar{S}}_i)^{\mu] \rho}
  - \frac{1}{2} ( S_i - {\bar{S}}_i )^{\mu \nu}
  \right. \nonumber \\
  & \left.
+\frac{q_{\rho}q_{\sigma}}{k_{i}\cdot q}
\Bigl[
\Bigl(
k_{i}^{[\mu}\bar{S}_{i}^{\nu]\rho}
+k_{i}^{[\nu}S_{i}^{\mu]\rho}
\Bigr)\partial_{i}^{\sigma}
+i
\Bigl(
S_{i}^{\rho [\mu} \bar{S}_{i}^{\nu]\sigma}
\Bigr)
\Bigr]
+
 q_\rho \Bigl[
 S_i^{\rho [\mu} \partial_i^{\nu]}
 +
 \bar{S}_i^{\rho [\nu} \partial_i^{\mu]}
 \Bigr]
 \right \}
 M_n \, ,
 \label{compactsubsub}
}
with 
\ea{
\tilde{A}_{\rm bosonic}^{\rho \mu \nu} =
&M_n
 \ast
 \sqrt{\frac{\a'}{2}} \sum_{i\neq j}
  \Bigg \{ 
  \frac{2i}{3} \left (\frac{\alpha'}{2}\right )^{3/2}\sum_{i\neq j\neq l}^n 
T^{\rho \mu \nu}(k_i, k_j,k_l)
  D_2\left (\frac{\zz{i}{m}}{\zz{i}{j}} \right ) 
\nonumber \\
&
+  
\left (\frac{\a'}{2} \right )
\frac{
T^{\rho \mu \nu}(\f_i \epsilon_i, k_i ,k_j)
}{\zz{i}{j}} 
+
\left (\frac{\a'}{2} \right )\sum_{l \neq i}
\frac{ 
T^{\rho \mu \nu}(\f_j \epsilon_j, k_i ,k_l)
}{\zz{i}{j}} \LN{i}{l}^2
\nonumber \\
&+
\frac{1}{2} \sqrt{\frac{\a'}{2}} \sum_{l \neq i}
\frac{
T^{\rho \mu \nu}(k_i, \f_j \epsilon_j ,\bar{\f}_l \bar{\epsilon}_l)
}{(\zz{i}{j})(\zbzb{i}{l})}
+
 \sqrt{\frac{\a'}{2}} \sum_{l\neq i}
 \frac{
 T^{\rho \mu \nu}(\f_i \epsilon_i, k_j ,{\f}_l {\epsilon}_l)
 }{(\zz{i}{j})(\zz{i}{l})} 
 \nonumber \\
&+
\frac{1}{3} \sum_{l\neq i, j}
\frac{
 T^{\rho \mu \nu}(\f_i \epsilon_i, \f_j \epsilon_j ,{\f}_l {\epsilon}_l)
}{(\zz{i}{j})(\zz{j}{l})(\zz{l}{i})}
 \Bigg\} 
 + \text{c.c.}
 \, .
 \label{compactsubsubA}
 }

It is tempting to think that, besides the dilogarithm term, all the other non-factorizing terms above may be reproducible in terms of gauge-invariant soft operators acting on the lower-point amplitude. We have investigated an exhaustive number of possibilities, and have not found any reduction as compared to the above expression.  
For instance, one could consider an operator involving $\epsilon_i^\alpha \partial_{\epsilon_i}^\beta \partial_i^\gamma$, where $\partial_{\epsilon_i}^\beta \equiv \partial/\partial \epsilon_{i\beta}$. This type of operator leads to the following type of terms
\ea{
&
\epsilon_i^\alpha \partial_{\epsilon_i}^\beta \partial_i^\gamma
M_n
=
M_n \ast
\sum_{j\neq i} \left (
    \alpha' k_j^\gamma \LN{i}{j} 
     - \sqrt{\frac{\alpha'}{2}} \frac{\te{\gamma}{j}}{\zz{i}{j}} -\sqrt{\frac{\alpha'}{2}} \frac{\tbe{\gamma}{j}}{\zbzb{i}{j}} \right )
     \nonumber \\
     &\times 
    \sum_{l\neq i} \left( 
\frac{\te{\alpha}{i}\te{\beta}{l}}{(\zz{i}{l})^2} + \sqrt{\frac{\alpha'}{2}} \frac{ \te{\alpha}{i} k_l^{\beta}}{\zz{i}{l}}
\right )
}
and is made gauge invariant by the combination:
\ea{
q_\rho \left(\epsilon_i^{[\mu} \partial_{\epsilon_i}^{\nu]} \partial_i^\rho
+
\epsilon_i^{\rho} \partial_{\epsilon_i}^{[\mu} \partial_i^{\nu]}
-
\epsilon_i^{[\mu} \partial_{\epsilon_i}^{\rho} \partial_i^{\nu]} \right ) M_n
}
This operator produces the same four types of terms as the last four lines in Eq.~\eqref{rest} plus two additional and different types of terms. However, among the four types that are similar it is only possible to match one of them, while the other three are different in their Koba-Nielsen structure. Therefore in total, while matching one line in Eq.~\eqref{rest}, five new gauge-invariant expressions are generated.

One instance, where one may introduce an additional soft operator, without elongating the expression, is 
\ea{
&\frac{\alpha'}{2}q_\rho\left  (k_i^{[\mu}   \epsilon_i^\rho \partial_{\epsilon_i}^{\nu]} + 
k_i^{[\nu}   \epsilon_i^{\mu]} \partial_{\epsilon_i}^{\rho} -
k_i^{\rho}   \epsilon_i^{[\mu} \partial_{\epsilon_i}^{\nu]} \right )M_n=
i\frac{\alpha' q_\rho}{2}\left  (
k_i^{[\mu}S_i^{\nu]\rho} + \frac{1}{2} k_i^{\rho} S_i^{\mu \nu}
\right )M_n
\nonumber \\
&=
M_n \ast \Bigg \{ 
\frac{\a'}{2} \sum_{j \neq i}
\frac{k_i^{[\mu}\te{\nu]}{j} (\teq{i})
+k_i^{[\nu} \te{\mu]}{i} (\teq{j}) -  (\qk{i}) \te{[\mu}{i}\te{\nu]}{j}
}{(\zz{i}{j})^2}
\nonumber \\
&+
\left (\frac{\a'}{2}\right )^{3/2} \sum_{j \neq i}
\frac{k_i^{[\mu}k_j^{\nu]} (\teq{i})
+k_i^{[\nu} \te{\mu]}{i} (\qk{j}) -  (\qk{i})  \te{[\mu}{i}k_j^{\nu]}
}{\zz{i}{j}}
\Bigg\}
 \nonumber \\
 &=
 M_n \ast \sqrt{\frac{\a'}{2}}
 q_\rho \Bigg \{ 
\sqrt{\frac{\a'}{2}} \sum_{j \neq i}
\frac{ T^{\rho \mu \nu}(\f_i \epsilon_i, k_i , \f_j\epsilon_j)
}{(\zz{i}{j})^2}
+
\left (\frac{\a'}{2}\right ) \sum_{j \neq i}
\frac{ T^{\rho \mu \nu}(\f_i \epsilon_i, k_i , k_j)
}{\zz{i}{j}}
\Bigg\}
\label{extrasofttheorem}
}
The last term matches the similar term in Eq.~\eqref{compactsubsubA}, but 
the first term is new. Therefore this operator effectively exchanges one type of 
term with another. We note that, although this operator is not exactly 
matching terms in Eq.~\eqref{compactsubsubA} one-to-one, it shows 
that terms of the form of the right-hand side above are terms of 
order $\alpha'$, because the operator explicitly carries such a factor.

\section{Soft scattering of $B_{\mu \nu}$ in superstrings }
\label{super}

For the derivation of the scattering amplitude
involving $n+1$ massless closed string states in superstrings we refer to Ref.~\cite{DMM3}.
As was shown in Ref.~\cite{DMM3}, we can take advantage of knowing the results in the bosonic string case, which were presented in the previous section.
This follows by realizing that the $n$-point tree-level scattering amplitudes, $\M_n$, of closed massless superstrings can generically be written as a convolution of a bosonic part, $M_n^b$, 
with a supersymmetric part, $M_n^s$, as follows:
\ea{
\M_n = M_n^b \ast M_n^s \, ,
}
The expressions for bosonic and supersymmetric parts of the $n$-point supersymmetric 
string amplitude are defined by: 
\seq{
M_n^b =  &\, \frac{8\pi}{\alpha'}\left (\frac{\kappa_D}{2\pi}\right )^{n-2}  
\int \frac{ \prod_{i=1}^{n} d^2 z_i}{dV_{abc} |z_1  - z_2|^2}  
{\prod_{i=1}^{2}  d \theta_i \theta_i  
\prod_{i=1}^{2} d \bar{\theta}_i \bar{\theta}_i}
\prod_{i=3}^{n} d \theta_i  
\prod_{i=1}^{n} d \varphi_i   
\prod_{i=3}^{n}   d {\bar{\theta}}_i  
\prod_{i=1}^n d {\bar{\varphi}}_i \\
&  \times  \prod_{i < j} | z_i - z_j |^{\alpha' k_i \cdot k_j}\, \exp \left[ \frac{1}{2} 
\sum_{i\neq j} \frac{C_i \cdot C_j}{(\zz{i}{j})^2} + \sqrt{\frac{\alpha'}{2}}
\sum_{i\neq j} \frac{C_i \cdot k_j}{\zz{i}{j}} + \text{c.c.} \right ] \, ,
}
\ea{
{M}_n^s = \exp \left [ - \frac{1}{2} \sum_{i\neq j} \frac{A_i \cdot A_j}{\zz{i}{j}} + 
\text{c.c.} \right] \, .\label{1.5}
}
where $\kappa_D$ is the $D$-dimensional Newton's constant,  $dV_{abc}$ 
is the volume of the M\"obius group, $z_i$ are the Koba-Nielsen variables,
$\varphi_i$ and $\theta_i$ are Grassmannian integration variables,
and  we have introduced the following \emph{superkinematical} quantities:
\ea{
 A_i^\mu = \varphi_i \epsilon_i^\mu  +\sqrt{\frac{\alpha'}{2}} \theta_i k_i^\mu
~~;
 ~~ C_i^\mu = \varphi_i \theta_i  \epsilon_i^\mu \, ,\label{1.4}
}
where $\epsilon_i^\mu$ and $k_i^\mu$ are respectively the holomorphic 
polarization vector and momentum of the state $i$, and $\alpha'$ is the string {Regge} slope.

Apart from the integration measure, $M_n^b$ is equivalent to the same amplitude in the bosonic string, given in Eq.~\eqref{nonsoftonly}; the integrands, in fact,  become equal if one makes the identification 
 $\theta_i \epsilon_i \to \epsilon_i$ and remembers
that, after this substitution,  $\epsilon_i$ becomes a Grassmann variable.
The difference between $M_n^b$ and the bosonic string amplitude Eq.~\eqref{nonsoftonly}, is only the presence in $M_n^b$ of the integrals over the Grassmann 
variables $\theta_i$, $\bar{\theta}_i$,  and the additional factor $\prod_{i=1}^2 
\theta_i \bar{\theta}_i/|z_1-z_2|^2$ coming from the correlator of 
the superghosts.

As in the case of the bosonic string,
it is also useful to factorize the superstring amplitude, at the integrand level, into a 
soft part $\S$ and a hard part as follows:
\ea{
\M_{n+1} = \M_n \ast \S
}
where $\M_n$ is the full superstring amplitude of $n$ closed massless states, 
and $\S$ is a function that when convoluted with the integral expression 
for $\M_n$ provides the additional soft state involved in the amplitude. 
The function $\S$ can further be decomposed into its bosonic part and 
supersymmetric part as follows:
\ea{
\S = S_b+S_s+\bar{S}_s \, ,
}
where $S_b$ is the purely bosonic part, given by: 
\seq{
 S_b= &\,\frac{\kappa_D}{2\pi} \int d^2 z \prod_{i=1}^{n}
 | z - z_i |^{ \alpha' q k_i} 
\, {\rm exp}\left [ -\sqrt{\frac{\alpha'}{2}}\frac{q \cdot C_i}{z-z_i}-  
\sqrt{\frac{\alpha'}{2}}\frac{q \cdot \bar{C}_i}{{\bar{z}}-{\bar{z}}_i} \right] 
 \\
& \times 
\left[ \sum_{i=1}^{n} \frac{\epsilon \cdot C_i}{(z-z_i)^2} +  
\sum_{i=1}^{n}\sqrt{\frac{\alpha'}{2}} \frac{\epsilon \cdot k_i}{z-z_i} \right] 
\left[ \sum_{i=1}^{n} \frac{\bar{\epsilon} \cdot 
{\bar{C}}_i}{({\bar{z}}-{\bar{z}}_i)^2} + \sum_{i=1}^{n}\sqrt{\frac{\alpha'}{2}} 
\frac{{\bar{\epsilon}} \cdot k_i}{{\bar{z}}-{\bar{z}}_i} \right] 
\, ,
}
which is simply equal to the similar expression in the bosonic string, given in Eq.~\eqref{last3lines}, after identifying $\theta_i \epsilon_i \to  \epsilon_i$ (whereby 
$\epsilon_i$ becomes a Grassmann variable). $S_s$ and $\bar{S}_s$ are the complex conjugates of each other and 
they provide the contributions from the additional supersymmetric {states}. 
They are given by
\seq{
\bar{S}_s = &\, \frac{\kappa_D}{2\pi} \int d^2 z \prod_{i=1}^{n}  
| z - z_i |^{\alpha' q\cdot  k_i} 
\, {\rm exp}\left [ -\sqrt{\frac{\alpha'}{2}}\frac{q \cdot C_i}{z-z_i}-  
\sqrt{\frac{\alpha'}{2}}\frac{q \cdot \bar{C}_i}{{\bar{z}}-{\bar{z}}_i} \right] \\
& \times \left[ \frac{1}{2}\sum_{i=1}^{n} \sqrt{\frac{\alpha'}{2}}
\frac{q \cdot A_i}{z-z_i} \sum_{j=1}^{n} \frac{\epsilon \cdot A_j}{z-z_j}
 \sum_{l=1}^{n}\sqrt{\frac{\alpha'}{2}} \frac{q \cdot
 {\bar{A}}_l}{{\bar{z}}-{\bar{z}}_l} \sum_{m=1}^{n} \frac{\bar{\epsilon} 
\cdot{\bar{A}}_m}{{\bar{z}}-{\bar{z}}_m} \right.
 \\
 &\left.
 + \left (\sum_{i=1}^{n} \frac{\epsilon \cdot C_i}{(z-z_i)^2} +  
\sum_{i=1}^{n}\sqrt{\frac{\alpha'}{2}} \frac{\epsilon \cdot k_i}{z-z_i} 
\right ) \sum_{j=1}^{n}\sqrt{\frac{\alpha'}{2}} \frac{q 
\cdot {\bar{A}}_j}{{\bar{z}}-{\bar{z}}_j} \sum_{l=1}^{n} \frac{\bar{\epsilon} 
\cdot{\bar{A}}_l}{{\bar{z}}-{\bar{z}}_l} 
  \right] \, ,
}
and  ${S}_s$ is given by  the complex conjugate of this expression,
where complex conjugation sends $z_i \to \bar{z}_i$, $\epsilon_i^\mu \to \bar{\epsilon}_i^\mu$, $\theta_i \to \bar{\theta}_i$, and $\varphi_i \to \bar{\varphi}_i$, while the momenta $k_i$ are left invariant. 
This decompositions is of course useful, since we already dealt with the bosonic integral $S_b$ in the previous section. The additional part coming from supersymmetry, $S_s + \bar{S}_s$, was furthermore computed through the order $q$ in Ref.~\cite{DMM3}.
Here we explicitly construct their functional form, when the soft state is an antisymmetric Kalb-Ramond field. The general result found in Ref.~\cite{DMM3} reads:
\ea{
&S_s+\bar{S}_s =
2 \kappa_D \epsilon_\mu \bar{\epsilon}_\nu  
 \sum_{i\neq j}  \Bigg \{
\frac{q_\rho}{(k_i\cdot q)} \frac{ \bar{A}_i^{[\rho}\bar{A}_j^{\nu]} 
k_i^\mu}{(\bar{z}_i-\bar{z}_j)}
 +
 q_\rho\left(\frac{\alpha'}{2}\right)^{\frac{3}{2}}
\frac{q\cdot k_j \bC_i^{[\rho,}k_i^{\nu]}}{\zbzb{i}{j}} 
\left (\frac{k_i^\mu}{q\cdot k_i} - \frac{k_j^\mu}{q\cdot k_j}\right ) 
 \nonumber \\
 &
 +q_\rho\sqrt{\frac{\alpha'}{2}}
\frac{\bA_{\{i ,}^\rho \bA_{j\} }^\nu}{\zbzb{i}{j}} \sum_{l\neq i} 
\Bigg [
\frac{q\cdot k_l}{q\cdot k_i} \left (\frac{C_i^\mu}{\zz{i}{l}} +
\sqrt{\frac{\alpha'}{2}} k_i^\mu \LN{i}{l}^2 \right )
 + 
\left (\frac{C_l^\mu}{\zz{i}{l}} - \sqrt{\frac{\alpha'}{2}}k_l^\mu \LN{i}{l}^2 \right )  
\Bigg ]
\nonumber \\
&
+q_\rho q_\sigma \Bigg[
\left (
\frac{1}{2}
A_{\{i,}^\sigma A_{j\}}^\mu -\sqrt{\frac{\alpha'}{2}} 
C_{\{i,}^\sigma k_{j\}}^\mu\right ) 
\sum_{l\neq i}\frac{2 \bA_{\{i ,}^\rho \bA_{l\} }^\nu}{q\cdot 
k_i (\zz{i}{j})(\zbzb{i}{l})} 
-\frac{\alpha'}{2}
\frac{\bC_i^{[\sigma,}k_i^{\nu]} \bC_j^\rho}{(\zbzb{i}{j})^2}
\left (\frac{k_j^\mu}{q\cdot k_j} - \frac{k_i^\mu}{q\cdot k_i}\right ) 
\nonumber \\
&
 -\sqrt{\frac{\alpha'}{2}} \sum_{l\neq i,j} \frac{k_i^\mu 
\left ( \bC_j^\sigma \bA_{\{ i,}^\rho \bA_{l\}}^\nu +\frac{1}{2} \bC_i^\sigma 
\bA_{\{ j,}^\rho \bA_{l\}}^\nu \right )}{q\cdot k_i (\zbzb{i}{j})(\zbzb{i}{l})}
- \sum_{l\neq i} \frac{2 C_{[i,}^\sigma C_{j]}^\mu \bA_{\{i ,}^\rho 
\bA_{l\} }^\nu}{q\cdot k_i(\zz{i}{j})^2 (\zbzb{i}{l})}  
\Bigg] \Bigg\} + \text{c.c.} + \Ord(q^2) \, ,
\label{SsbarSs}
}
where the brackets and curly-brackets in the indices denotes 
commutation and anticommutation of the indices, e.g.:
\seq{
 C_i^{[\rho ,} k_i^{\nu]} &\equiv \frac{1}{2} \left (C_i^\rho k_i^\nu - C_i^\nu k_i^\rho \right )\\
 A_{\{i}^\mu A_{j\}}^\nu &\equiv \frac{1}{2} \left ( A_i^\mu A_j^\nu + A_j^\mu A_i^\nu\right ) \, .
 }
 These definitions differ by a factor of two with the ones 
in Ref.~\cite{DMM3}, 
{where} also appropriate factors of two have
 been introduced in Eq.~\eqref{SsbarSs}.

We must now project out the antisymmetric part, in $\epsilon_\mu \bar{\epsilon}_\nu$, to obtain the expression for the Kalb-Ramond field.
At order $q^0$ we find:
\ea{
&S_s+\bar{S}_s\Big |_B^{\mu \nu} =
 \sum_{i\neq j} 
\frac{q_\rho}{(k_i\cdot q)} 
\left(
\frac{ 
{A}_{\{i}^{\rho}{A}_{j\}}^{[\mu}
k_i^{\nu]}}{({z}_i-{z}_j)}
+
\frac{ 
\bar{A}_{\{i}^{\rho}\bar{A}_{j\}}^{[\nu}
k_i^{\mu]}}{(\bar{z}_i-\bar{z}_j)}
\right) + \Ord(q)
\label{antisubleading}
}

At the next order, it can be checked that the antisymmetric part can be written as:
\ea{
&S_s+\bar{S}_s |_{\Ord(q)}=
 \kappa_D  \epsilon_{\mu \nu}^B
 \sum_{i\neq j}^n \Bigg \{
 q_\rho 
\left(\frac{\alpha'}{2}\right)^{\frac{3}{2}}
\frac{
T^{\rho \mu \nu} (\bC_i, k_i , k_j)
}{\zbzb{i}{j}} 
+
\sqrt{\frac{\alpha'}{2}}\frac{q_\rho q_\sigma}{q\cdot k_i} 
\frac{\bA_i^\sigma \bA_i^\nu \bC_j^\rho k_i^\mu}{(\zbzb{i}{j})^2}
\nonumber \\
&
+
 \left ({\frac{\alpha'}{2}}\right )
 \sum_{l\neq i}
\frac{ 4 q_\rho q_\sigma}{\q \cdot k_i} 
 \frac{\bA_{\{i ,}^\rho \bA_{j\} }^{\nu}
k_i^{[\mu} k_l^{\sigma]} 
 }{\zbzb{i}{j}} \LN{i}{l}^2
+ \sqrt{\frac{\alpha'}{2}}\sum_{l\neq i}
\frac{ 8 q_\rho q_\sigma}{\q \cdot k_i}
\frac{\bA_{\{i ,}^\rho \bA_{j\}}^\nu C_{\{i}^{[\mu}k_{l\}}^{\sigma]}  }{(\zbzb{i}{j})(\zz{i}{l})}
 \nonumber \\
&
+\sum_{l\neq i}
 \frac{2q_\rho q_\sigma }{q\cdot k_i}
\frac{A_{\{i,}^\sigma A_{j\}}^\mu \bA_{\{i ,}^\rho \bA_{l\} }^\nu}{ (\zz{i}{j})(\zbzb{i}{l})} 
-\sum_{l\neq i}
\frac{4 q_\rho q_\sigma}{q \cdot k_i} \frac{ C_{[i,}^\sigma C_{j]}^\mu \bA_{\{i ,}^\rho 
\bA_{l\} }^\nu}{(\zz{i}{j})^2 (\zbzb{i}{l})} 
\nonumber \\
&
-
\sqrt{\frac{\alpha'}{2}}\frac{q_\rho q_\sigma}{q\cdot k_i} 
\sum_{l\neq i}
\frac{\bC_i^\sigma k_i^\mu  
\bA_{\{ j,}^\rho \bA_{l\}}^\nu }{ (\zbzb{i}{j})(\zbzb{i}{l})}
 -
 \sqrt{\frac{\alpha'}{2}}\frac{q_\rho q_\sigma}{q\cdot k_i}  \sum_{l\neq i,j} \frac{
 2\bC_j^\sigma k_i^\mu  \bA_{\{ i,}^\rho \bA_{l\}}^\nu}{ (\zbzb{i}{j})(\zbzb{i}{l})}
\Bigg \}
 + \text{c.c.}
 \label{antisubsub}
}
where $T^{\rho \mu \nu}$ was defined in Eq.~\eqref{Atensor}.
It can be checked that the term involving $T^{\rho \mu \nu}$, in fact, cancels out with the similar term coming from the bosonic part in Eq.~\eqref{compactsubsubA}.

\subsection{The soft theorem}
As remarked in Sec.~\ref{bosonicsubleading}, 
the soft theorem can be equivalently written 
as in Eq.~\eqref{generalorderq0operator}.
Since the soft theorem operator is a linear operator,
and since we know {that}
it reproduces the bosonic part, when acting on $M_n^b$, we only need to show that it reproduces the supersymmetric part, given in Eq.~\eqref{antisubleading}, when acting on $M_n^s$, i.e.
\ea{
-i \epsilon_{\mu\nu}^B \sum_{i=1}^n
\left[\frac{q_\rho \bar{k}_i^\nu(L_i+S_i)^{\mu\rho}}{qk_i}
+\frac{q_\rho k_i^\mu(\bar{L}_i+\bar{S}_i)^{\nu\rho}}{qk_i}\right]
M_n^s(k_i,\epsilon_i;\bar{k}_i\bar{\epsilon}_i)\Big|_{k=\bar{k}}
= 
M_n^s \ast (S_s + \bar{S}_s)
}
which is easily checked by noticing the relations:
\seq{
&(L_i+S_i)^{\mu \rho} A_j^\sigma = i \delta_{ij} \left (\eta^{\sigma \rho} A_i^\mu - 
\eta^{\sigma \mu} A_i^\rho \right ) \, , \quad
(\bar{L}_i+\bar{S}_i)_i^{\mu \rho} \bA_j^\sigma = i \delta_{ij} \left (\eta^{\sigma \rho} 
\bA_i^\mu - \eta^{\sigma \mu} \bA_i^\rho \right ) \, ,
\\
&(L_i+S_i)^{\mu \rho} C_j^\sigma = i \delta_{ij} \left (\eta^{\sigma \rho} C_i^\mu - 
\eta^{\sigma \mu} C_i^\rho \right ) \, , \quad
(\bar{L}_i+\bar{S}_i)^{\mu \rho} \bC_j^\sigma = i \delta_{ij} \left (\eta^{\sigma \rho} 
\bC_i^\mu - \eta^{\sigma \mu} \bC_i^\rho \right ) \, .
\label{JC}
}
{where the quantities $A$ and $C$ are defined in Eq.~(\ref{1.4}).} 

\subsection{The order $q$ soft behavior}
Let us now consider the order $q$ soft operator, 
which up to the totally antisymmetric tensor according to Eq.~\eqref{compactsoftoperator}, reads:
\ea{
\hat{S}_B^{(1)} = \epsilon_{q\, \mu\nu}^B \sum_{i=1}^{n}
\frac{q_{\rho}q_{\sigma}}{k_{i}\cdot q}
J_{i}^{\rho [\mu} \bar{J}_{i}^{\nu]\sigma}
}
The action of this operator on $\M_n$ can be decomposed as follows
\ea{
\hat{S}_B^{(1)} \M_n = &(\hat{S}_B^{(1)} M_n^b)\ast M_n^s + M_n^b\ast( \hat{S}_B^{(1)} M_n^s)
\nonumber \\ 
&
+
\epsilon_{q\, \mu\nu}^B \sum_{i=1}^{n}
\frac{q_{\rho}q_{\sigma}}{k_{i}\cdot q}
\left [
(J_{i}^{\rho [\mu} M_n^b)\ast(  \bar{J}_{i}^{\nu]\sigma} M_n^s)
+
(  \bar{J}_{i}^{\sigma [\nu} M_n^b)\ast (J_{i}^{ \mu ]\rho } M_n^s)
\right ]
\label{DecomposeSoftB}
}
The action $\hat{S}_B^{(1)} M_n^b$ is equivalent to the one studied in the bosonic string. 
We need to analyze the remaining terms.
Let us notice that 
\seq{
&J_i^{\mu \nu} A_i^\rho = 2 i A_i^{[\mu}\eta^{ \nu] \rho}
\, , \
J_i^{\mu \nu} C_i^\rho = 2 i C_i^{[\mu}\eta^{ \nu] \rho} \\
&
J_i^{\mu \nu} \bA_i^\rho = 2 i \sqrt {\frac{\a'}{2}} \bar{\theta}_i k_i^{[\mu}\eta^{ \nu] \rho}
\, , \
J_i^{\mu \nu} \bC_i^\rho = 0
}
and likewise for the antiholomorphic counterpart.
We find that
\ea{
J_{i}^{\rho \mu} \bar{J}_{i}^{\nu\sigma} M_n^s \sim&
- 4 M_n^s \ast 
\sum_{j\neq i}\sum_{l\neq i}
\left [  \frac{A_i^{[\rho} A_j^{\mu ]}\bA_i^{[\nu} \bA_l^{\sigma ]}}{(\zz{i}{j})(\zbzb{i}{l})} 
+
\frac{\a'}{2} \frac{\bar{\theta}_i k_i^{[\rho} \bA_j^{\mu]}{\theta}_i k_i^{[\nu} A_l^{\sigma]}}{(\zbzb{i}{j})(\zz{i}{l})}
\right. \nonumber \\
&\left. +\sqrt{\frac{\a'}{2}}
\frac{\bar{\theta}_i k_i^{[\rho} \bA_j^{\mu]} \bA_i^{[\nu} \bA_l^{\sigma ]}}{(\zbzb{i}{j})(\zbzb{i}{l})}
+\sqrt{\frac{\a'}{2}}
\frac{{\theta}_i k_i^{[\sigma} A_j^{\nu]} A_i^{[\mu} A_l^{\rho ]}}{(\zz{i}{j})(\zz{i}{l})}
\right ] \, .
}
The first term above reproduces exactly a similar term of the explicit result of Eq.~\eqref{antisubsub}.
The terms with one $\theta_i$ can be reduces as follows, using that $\theta_i A_i^\mu =- C_i^\mu$:
\ea{
-4\sqrt{\frac{\a'}{2}}
\frac{\bar{\theta}_i k_i^{[\rho} \bA_j^{\mu]} \bA_i^{[\nu} \bA_l^{\sigma ]}}{(\zbzb{i}{j})(\zbzb{i}{l})}
=&
-2\sqrt{\frac{\a'}{2}}
\frac{\bar{\theta}_i k_i^{\rho} \bA_j^{\mu} \bA_i^{[\nu} \bA_l^{\sigma]}
}{(\zbzb{i}{j})(\zbzb{i}{l})}
-\sqrt{\frac{\a'}{2}}
\frac{\bC_i^\sigma k_i^\mu \bA_{\{j}^\rho \bA_{l\}}^\nu}{(\zbzb{i}{j})(\zbzb{i}{l})}
\label{JJonetheta}
} 
where we also used that $q_\rho q_\sigma \frac{\bA_j^\rho \bA_l^\sigma}{ (\zbzb{i}{j})(\zbzb{i}{l})}= 0$, because the denominator and the numerator have different parity.
 The other term with one $\theta_i$ is the complex conjugate of this expression. 
 The second term above also reproduces a similar term of the explicit result of Eq.~\eqref{antisubsub}.

It is finally useful to simplify the terms with $\theta_i \bar{\theta}_i$, which due to 
$\epsilon_{\mu \nu}^B k_i^\mu k_i^\nu = 0$, can be reduced to:
\ea{
-4\frac{q_\rho q_\sigma}{q\cdot k_i} \frac{\a'}{2} \frac{\bar{\theta}_i k_i^{[\rho} \bA_j^{\mu]}{\theta}_i k_i^{[\nu} A_l^{\sigma]}}{(\zbzb{i}{j})(\zz{i}{l})}
&=
-q_\rho \frac{\a'}{2} \frac{\theta_i \bar{\theta}_i ( 
k_i^\nu \bA_j^\mu A_l^\rho
- k_i^\rho \bA_j^\mu A_l^\nu
+ k_i^\mu \bA_j^\rho A_l^\nu
)
 }{(\zbzb{i}{j})(\zz{i}{l})}
 \nonumber \\
 &=
- q_\rho \frac{\a'}{2} \frac{\theta_i \bar{\theta}_i 
\, T^{\rho \mu \nu}(k_i, A_j, \bA_l)
 }{(\zbzb{i}{l})(\zz{i}{j})} \, ,
}
showing that this term is totally gauge invariant and local in $q$.

Next, consider the second line of Eq.~\eqref{DecomposeSoftB}.
It is easy to derive the following action:
\ea{
(J_{i}^{\rho \mu} M_n^b)&\ast(  \bar{J}_{i}^{\nu\sigma} M_n^s)
= 
\nonumber \\
&
 4 \M_n \ast
\sum_{j\neq i}
\left [
\a' k_i^{[\rho}k_j^{\mu]} \LN{i}{j} + \frac{C_i^{[\rho} C_j^{\mu]}}{(\zz{i}{j})^2} 
+\sqrt{\frac{\a'}{2}} \left( \frac{2 C_{\{i}^{[\rho} k_{j\}}^{\mu]}}{\zz{i}{j}} + \frac{\bC_j^{[\rho} k_i^{\mu]}}{\zbzb{i}{j}} \right )\right ] \nonumber \\
&\times 
\sum_{l\neq i}\left [  \frac{\bA_i^{[\nu} \bA_l^{\sigma ]}}{\zbzb{i}{l}} + \sqrt{\frac{\a'}{2}} \frac{{\theta}_i k_i^{[\nu} A_l^{\sigma]}}{\zz{i}{l}} \right ] 
}
The antiholomorphic version of this is simply recovered by complex conjugation, i.e.
\ea{
(J_{i}^{\sigma \nu} M_n^b)&\ast(  \bar{J}_{i}^{\mu\rho} M_n^s)
= 
\nonumber \\
&
 4 \M_n \ast
\sum_{j\neq i}
\left [
\a' k_i^{[\sigma}k_j^{\nu]} \LN{i}{j} + \frac{\bC_i^{[\sigma} \bC_j^{\nu]}}{(\zbzb{i}{j})^2} 
+\sqrt{\frac{\a'}{2}} \left( \frac{2 \bC_{\{i}^{[\sigma} k_{j\}}^{\nu]}}{\zbzb{i}{j}} + \frac{C_j^{[\sigma} k_i^{\nu]}}{\zz{i}{j}} \right )\right ] \nonumber \\
&\times 
\sum_{l\neq i}\left [  \frac{A_i^{[\mu} A_l^{\rho ]}}{\zz{i}{l}} + \sqrt{\frac{\a'}{2}} \frac{\bar{\theta}_i k_i^{[\mu} \bA_l^{\rho]}}{\zbzb{i}{l}} \right ] 
}
Let us call the terms in first square bracket L1, L2, L3 and L4, and the terms in the second square bracket R1 and R2.

It is fairly easy to see that the multiplication of $(L1 + L2 + L3) \times R1$ 
produces terms that can directly be matched with terms in Eq.~\eqref{antisubsub}.
The term $L4\times R1$ can be written as:
\ea{
4 \sqrt{\frac{\a'}{2}} \frac{C_j^{[\sigma} k_i^{\nu]}}{(\zz{i}{j})}
 \frac{A_i^{[\mu} A_l^{\rho ]}}{(\zz{i}{l})}
=
2 \sqrt{\frac{\a'}{2}} \frac{C_j^{\sigma} k_i^{\nu}}{(\zz{i}{j})}
 \frac{A_{\{i}^{\mu} A_{l\}}^{\rho}}{(\zz{i}{l})}
-2 \sqrt{\frac{\a'}{2}} \frac{(-\theta_j A_j^{\nu}) k_i^{\sigma}}{(\zz{i}{j})}
 \frac{A_{\{i}^{\mu} A_{l\}}^{\rho}}{(\zz{i}{l})}
 \label{L4R1}
}
The first term on the right hand side matches a similar term in Eq.~\eqref{antisubsub},
while the second term, which was rewritten using $C_j = - \theta_j A_j$, remains unmatched.

All the terms multiplying R2 above also remain unmatched, however, they all 
simplify to a local, totally antisymmetric {expression}, due 
to $\theta_i C_i = 0$ 
and $k_i^\mu k_i^\nu =0$. Specifically we find:
\ea{
&  \frac{4 q_\rho q_\sigma}{q\cdot k_i} \left [
\a' k_i^{[\sigma}k_j^{\nu]} \LN{i}{j} + \frac{\bC_i^{[\sigma} \bC_j^{\nu]}}{(\zbzb{i}{j})^2} 
+\sqrt{\frac{\a'}{2}} \left( \frac{2 \bC_{\{i}^{[\sigma} k_{j\}}^{\nu]}}{\zbzb{i}{j}} + \frac{C_j^{[\sigma} k_i^{\nu]}}{\zz{i}{j}} \right )\right ]
\times 
 \sqrt{\frac{\a'}{2}} \frac{\bar{\theta}_i k_i^{[\mu} \bA_l^{\rho]}}{\zbzb{i}{l}}
\nonumber \\
&=
\frac{a'}{2}q_\rho \bar{\theta}_i
\Bigg [
\sqrt{\frac{a'}{2}}\frac{ k_j^\nu k_i^\mu \bA_l^\rho - k_j^\nu k_i^\rho \bA_l^\mu+k_i^\nu k_j^\rho \bA_l^\mu}{\zbzb{i}{l}} \LN{i}{j}^2
\nonumber \\
& \quad
+\frac{- \bC_j^\rho k_i^\nu \bA_l^\mu - \bC_j^\nu k_i^\mu \bA_l^\rho +\bC_j^\nu k_i^\rho \bA_l^\mu}{(\zbzb{i}{j})(\zbzb{i}{l})}
+\frac{- C_j^\rho k_i^\nu \bA_l^\mu - C_j^\nu k_i^\mu \bA_l^\rho +C_j^\nu k_i^\rho \bA_l^\mu}{(\zz{i}{j})(\zbzb{i}{l})}
\Bigg ]
\nonumber \\
&=
\frac{a'}{2}q_\rho \bar{\theta}_i
\Bigg [
\sqrt{\frac{a'}{2}}\frac{T^{\mu \nu \rho}(k_i, k_j, \bA_l)}{\zbzb{i}{l}} \LN{i}{j}^2
-\frac{T^{\mu \nu \rho}(k_i, \bC_j, \bA_l)}{(\zbzb{i}{j})(\zbzb{i}{l})}
-\frac{T^{\mu \nu \rho}(k_i, C_j, \bA_l)}{(\zz{i}{j})(\zbzb{i}{l})}
\Bigg ]
}

Up to the totally antisymmetric terms local in $q$, there remains three terms 
that {need} to be rewritten;
one is an unmatched term in the explicit result of Eq.~\eqref{antisubsub},
and the other two are the unmatched terms of Eq.~\eqref{JJonetheta} and \eqref{L4R1}.
Subtracting the latter two from the former, we find:
\ea{
 \sqrt{\frac{\a'}{2}} q_\rho \left [ \frac{ \bA_i^{\mu} \bA_i^{\nu}  \bC_j^\rho }{2(\zbzb{i}{j})^2} 
+2\sum_{l\neq i}
\frac{\bar{\theta}_i \bA_j^{\mu} \bA_i^{[\nu} \bA_l^{\rho]}
}{(\zbzb{i}{j})(\zbzb{i}{l})}
-2\sum_{l\neq i} \frac{\bar{\theta}_j \bA_j^{\mu}}{(\zbzb{i}{j})}
 \frac{\bA_{\{i}^{\nu} \bA_{l\}}^{\rho}}{(\zbzb{i}{l})}
\right ]
+ \text{c.c}
}
Let us consider the expression when $l=j$. Then the last term vanishes, since $\theta_j A_j A_j = 0$, and we are left with (up to the prefactor and the complex conjugates):
\ea{
&\frac{1}{(\zbzb{i}{j})^2} 
\left [
\frac{1}{2}
\bA_i^{\mu} \bA_i^{\nu}  \bC_j^\rho 
+2\bar{\theta}_i \bA_j^{\mu} \bA_i^{[\nu} \bA_j^{\rho]}
\right ]
 =
\frac{1}{(\zbzb{i}{j})^2} 
\left [
\bC_j^{\nu}\bA_i^{\mu} \bA_i^{\rho}-
\frac{1}{2}\bC_j^{\rho}\bA_i^{\mu} \bA_i^{\nu}
\right ]
\nonumber\\
&
=
\frac{1}{(\zbzb{i}{j})^2} 
\sqrt{\frac{\a'}{2}}
\left [
2\bC_j^{\nu}\bC_i^{[\mu} k_i^{\rho]}-
\bC_j^{\rho}\bC_i^{\mu} k_i^{\nu}
\right ]
=
-\sqrt{\frac{\a'}{2}}\frac{T^{\mu \nu \rho}(\bC_i, k_i, \bC_j)}{(\zbzb{i}{j})^2} 
}
where in the end we also used antisymmetry of the $\mu \nu$ indices.
This shows that the terms above for  $l = j$ form a gauge invariant combination.
Next we consider the $l\neq j$ terms (again suppressing the overall prefactors):
\ea{
&2\sum_{l\neq i, j}\left [
\frac{\bar{\theta}_i \bA_j^{\mu} \bA_i^{[\nu} \bA_l^{\rho]}
}{(\zbzb{i}{j})(\zbzb{i}{l})}
- \frac{\bar{\theta}_j \bA_j^{\mu}}{(\zbzb{i}{j})}
 \frac{\bA_{\{i}^{\nu} \bA_{l\}}^{\rho}}{(\zbzb{i}{l})}
\right ]
\nonumber \\
&
=
2\sum_{l\neq i, j}\bar{\theta}_i \left [
\frac{\bA_j^{\mu} \bA_i^{[\nu} \bA_l^{\rho]}
}{(\zbzb{i}{j})(\zbzb{i}{l})}
+
\bA_i^{\mu}\bA_{\{j}^{\nu} \bA_{l\}}^{\rho}
 \left(
\frac{1 }{2(\zbzb{i}{j})(\zbzb{j}{l})}
+
\frac{1 }{2(\zbzb{i}{l})(\zbzb{l}{j})}
\right )
\right ]
\nonumber \\
&
=
 \sum_{l\neq i, j}\bar{\theta}_i \left [
 \frac{\bA_i^{\mu} \bA_j^{\nu} \bA_l^{\rho}
 +\bA_i^{\rho} \bA_j^{\mu}  \bA_l^{\nu}
 +
 \bA_i^{\nu}\bA_{j}^{\rho} \bA_{l}^{\mu}
 }{(\zbzb{i}{j})(\zbzb{i}{l})}
 \right ]
=
 \sum_{l\neq i, j}\bar{\theta}_i
 \frac{
T^{\mu \nu\rho}(
\bA_i, \bA_j,\bA_l)
 }{(\zbzb{i}{j})(\zbzb{i}{l})}
}
where for the first equality we manipulated the summation indices $i, j, l$, for the second equality we manipulated the expression using both $\mu \nu$ antisymmetry and $j,l$ symmetry. Again the final expression is, consistently, a totally antisymmetric, gauge invariant term.

We are now in a position to write our entire result in terms of the totally antisymmetric tensor. Adding also the bosonic part, derived in Eq.~\eqref{compactsubsub} with $\epsilon_i \to \theta_i \epsilon_i$, we find:
\ea{
&
\M_n \ast (S_b+S_s + \bar{S}_s)|_{\Ord(q)}
-\epsilon_{q\, \mu\nu}^B \sum_{i=1}^{n}
\frac{q_{\rho}q_{\sigma}}{k_{i}\cdot q}
J_{i}^{\rho [\mu} \bar{J}_{i}^{\nu]\sigma}
\M_n 
=
  q_\rho \tilde{A}_{\rm super}^{\rho \mu \nu} \, ,
}
with
\ea{
\tilde{A}_{\rm super}^{\rho \mu \nu}
=
&\M_n \ast 
 \kappa_D  \epsilon_{\mu \nu}^B
 \sqrt{\frac{\a'}{2}}
\sum_{i\neq j}
  \Bigg \{ 
-\sqrt{\frac{\a'}{2}}\frac{T^{\mu \nu \rho}(\bC_i, k_i, \bC_j)}{(\zbzb{i}{j})^2} 
\nonumber \\
&
+
\left (\frac{\a'}{2} \right )\sum_{l \neq i}
\frac{
T^{\rho \mu \nu}(C_j-\theta_i A_j, k_i ,k_l)
}{\zz{i}{j}} \LN{i}{l}^2
\nonumber \\
&+
\frac{1}{2} \sqrt{\frac{\a'}{2}} \sum_{l \neq i}
\frac{
T^{\rho \mu \nu}(k_i, C_j ,\bC_l)
-
2 \theta_i T^{\rho \mu \nu}(k_i, A_j, \bC_l)
-
\theta_i \bar{\theta}_i
T^{\rho \mu \nu}(k_i, A_j, \bA_l)
}{(\zz{i}{j})(\zbzb{i}{l})}
\nonumber \\
&
+
 \sqrt{\frac{\a'}{2}} \sum_{l\neq i}
 \frac{
 T^{\rho \mu \nu}(C_i, k_j ,C_l)
 +
 \theta_i T^{\rho \mu \nu }(k_i, C_j, A_l)
 }{(\zz{i}{j})(\zz{i}{l})} 
  + \sum_{l\neq i, j}{\theta}_i
 \frac{
T^{\mu \nu\rho}(
A_i, A_j,A_l)
 }{(\zz{i}{j})(\zz{i}{l})}
 \nonumber \\
&+
\frac{1}{3} \sum_{l\neq i, j}
\frac{
 T^{\rho \mu \nu}(C_i, C_j ,C_l)
}{(\zz{i}{j})(\zz{j}{l})(\zz{l}{i})}
 \nonumber \\
&
 +
\frac{2i}{3} \left (\frac{\alpha'}{2}\right )^{3/2}\sum_{l \neq i, j}^n 
T^{\rho \mu \nu}(k_i, k_j,k_l)
  D_2\left (\frac{\zz{i}{m}}{\zz{i}{j}} \right ) 
\Bigg \}+ \text{c.c}
}
As in the bosonic string, we conclude, based on the dilogarithmic  terms, 
that the expression above cannot be expressed as an operator acting on 
the lower point amplitude. Especially, supersymmetry does not provide 
enough simplifications for this to happen. We have not attempted to 
study the field theory implications of these results, where 
especially the dilogarithmic terms should vanish, and the 
conclusions about the order $q$ factorization there remains an open question.

\vfill

\section{Conclusion}
\label{conclusion}

We have shown using gauge invariance that the soft behavior of the 
antisymmetric B-field is fixed at the  order $q^0$ in the soft momentum in
amplitudes
involving gravitons, dilatons and other B-fields. We have furthermore 
explained why gauge invariance cannot fix completely its  
soft behavior {at order $q$}, in contrast to the case of a soft graviton or dilaton. By 
using the leading soft theorem, it is nevertheless possible to explicitly 
decompose the amplitude into two separately gauge invariant parts 
to all orders in the soft momentum.

The new soft theorem provides, together with the soft theorems 
through the same order for the graviton and dilaton, the basis for
 the unification of the three soft theorems, which we have offered 
in Eq.~\eqref{unified}. This universal expression is a step towards 
understanding the interplay between the infrared behaviors of
Yang-Mills theory and gravity as Yang-Mills 
squared theory~\cite{Cheung:2017ems}.

We have explicitly checked the new soft theorem of the B-field in 
the bosonic string as well as in superstrings, and we have furthermore
 computed the soft behavior through order $q$
in both theories, expressed in terms 
of a convoluted integral of a hard and a soft part. Based on the structure 
of the soft integrand we conclude in both theories that the 
soft behavior at order $q$ cannot be factorized in form of a soft theorem. As regards 
the field theory limit of these expression, the conclusion remains an 
open question and should be further studied.

\subsection*{Acknowledgments} \vspace{-3mm}
We owe special thanks to Josh Nohle for collaborating with us in the early 
stages of this work.

\end{document}